\documentclass{aa}
\usepackage{txfonts}
\usepackage[authoryear]{natbib}
\usepackage{natbib}
\usepackage{amssymb}
\bibpunct{(}{)}{;}{a}{}{,}
\usepackage{epsf}
\usepackage{graphicx}
\usepackage{epsfig}
\usepackage{graphics}
\usepackage{subfigure}
\usepackage[english]{babel}
\usepackage{rotating}
\usepackage{color}

\newcommand{\teff}{$T_{\rm{eff}}$}
\newcommand{\logg}{$\log g$}
\newcommand{\lL}{\ifmmode \log \frac{L}{L_{\sun}} \else $\log \frac{L}{L_{\sun
}}$\fi}
\newcommand{\mdot}{$\dot{M}$}
\newcommand{\myr}{M$_{\sun}$ yr$^{-1}$}

\newcommand{\vinf}{$v_{\infty}$}
\newcommand{\vturb}{$v_{\rm turb}$}

\newcommand{\kms}{km~s$^{-1}$}
\newcommand{\msun}{M$_{\sun}$}

\newcommand{\ciii}{\ion{C}{iii} 5696}

\newcommand{\ltu}{2s3p~$^3$P$^{\scriptsize \rm o}$}
\newcommand{\ltl}{2s3s~$^3$S}
\newcommand{\lsl}{2s3p~$^1$P$^{\scriptsize \rm o}$}
\newcommand{\lsu}{2s3d~$^1$D}

\begin{document}

\title{On the formation of \ion{C}{iii} 4647-50-51 and \ion{C}{iii} 5696 in O star atmospheres}
\author{F. Martins\inst{1}
\and D.J. Hillier\inst{2}
}
\institute{LUPM--UMR 5299, CNRS \& Universit\'e Montpellier II, Place Eug\`ene Bataillon, F-34095, Montpellier Cedex 05, France \\
           \email{fabrice.martins@univ-montp2.fr}
\and
             Department of Physics and Astronomy, University of Pittsburgh, 3941 O'Hara Street, Pitts
burgh, PA 15260, USA 
}

\date{Received / }

\abstract
{Surface abundances of OB stars provide important information on  mixing processes in stellar interiors and on evolutionary status. Performing abundance determination from selected C, N and O lines in the optical spectra of OB stars is thus necessary to understand the evolution of massive stars. }
{In this study, we investigate the formation of \ion{C}{iii} 4647--51--50 and \ciii\ in the atmosphere of O stars to see if they can be reliably used for abundance determinations.}
{We use atmosphere models computed with the code CMFGEN. The key physical ingredients explaining the formation of the \ion{C}{iii} lines are extracted from comparisons of models with different stellar parameters and through examining rates controlling the level populations.}
{The strength of \ciii\ critically depends on UV \ion{C}{iii} lines at 386, 574 and 884 \AA. These lines control the \ciii\ upper and lower level population. \ion{C}{iii} 884 plays a key role in late O stars where it drains the lower level of \ciii. \ion{C}{iii} 386 and \ion{C}{iii} 574 are more important at early spectral types. The overlap of these UV lines with \ion{Fe}{iv} 386.262, \ion{Fe}{iv} 574.232 and \ion{S}{v} 884.531 influences the radiative transfer at 386, 574 and 884 \AA, and consequently affects the strength of \ciii. \ion{C}{iii} 4650 is mainly controlled by the \ion{C}{iii} 538 line which acts as a drain on its lower level. \ion{Fe}{iv} 538.057 interacts with \ion{C}{iii} 538 and has an impact on the \ion{C}{iii} 4650 profile. 
Low temperature dielectronic recombinations have a negligible effect on the line profiles.
The main effect of the wind on the \ion{C}{iii} lines is an increase of the emission as mass loss rate increases, due to a shift of the formation region towards high velocity. Such a shift also implies desaturation of \ion{C}{iii} 386,  \ion{C}{iii} 538,  \ion{C}{iii} 574 and \ion{C}{iii} 884 which in turn modify the \ion{C}{iii} 4650 and \ciii\ line profiles.  
Metallicity effects are complex and result from an interplay between ionization, metallic line interaction with UV \ion{C}{iii} lines, and wind effects. }
{Given our current understanding of the stellar and wind properties of O stars, and in view of the present results, the determination of accurate carbon abundances from  \ion{C}{iii} 4647--51--50 and \ciii\ is an extremely challenging task. Uncertainties lower than a factor of two on C/H determinations based only on these two sets of lines should be regarder as highly doubtful. Our results provide a possible explanation of the variability of \ion{C}{iii} 4650 in Of?p stars.}


\authorrunning{Martins \& Hillier}
\titlerunning{\ion{C}{iii} line formation in O stars}

\maketitle

%%%%%%%%%%%%%%%%%%%%%%%%%%%%%%%%%%%%%%%%%%%%%%%%%%%%%%%%%%%%%%%%%%%%%%%%%%%%%%%%%%%%%%%%%%%%%%%%%%%%%%%%%%%%%%%%%%%%%%%%%%%%%%%
%%%%%%%%%%%%%%%%%%%%%%%%%%%%%%%%%%%%%%%%%%%%%%%%%%%%%%%%%%%%%%%%%%%%%%%%%%%%%%%%%%%%%%%%%%%%%%%%%%%%%%%%%%%%%%%%%%%%%%%%%%%%%%%
\section{Introduction}
\label{s_intro}

Massive stars turn hydrogen into helium through the CNO cycle in their core. In this process, nitrogen is produced at the expense of carbon and oxygen. Mixing processes subsequently transport the products of nucleosynthesis from the core to the outer layers and surface of the star. The spectrum of a massive star thus contains the imprints of nuclear reactions taking place in the interior of the star.
The older a star, the larger its nitrogen content and the lower its carbon abundance (provided the star is of type O and not yet an evolved object such as Wolf-Rayet (W-R), red supergiant or Luminous Blue Variable). The surface chemical composition depends on the star's rotation rate, metallicity, and mass. A fast rotation, for example, favors the transport of chemical species and thus quicker enrichment of the photosphere by material synthesized in the core \citep{mm00}. Low metallicity stars burn hydrogen more efficiently, as does a 60 \msun\ stars compared to a 20 \msun\ star. Chemical surface abundances are thus a key to understand the physical processes controlling the evolution of massive stars. It was surface abundances that provided key clues that W-R stars were evolved \citep[e.g.,][]{S73_Habund,SW82_CNO}.  Recent results \citep{hunter08,przy10,brott11a,martins12} have shown that strong constraints on the physics of massive stars could be obtained by spectroscopic analysis and abundance studies.

Nevertheless, the determination of surface abundances of O stars is challenging. Non--LTE atmosphere models including line--blanketing and winds are required, and there is a rarity of metallic lines in O star spectra. The UV range contains  a few strong resonance lines (\ion{C}{iii} 1176, \ion{N}{v} 1240, \ion{C}{iv} 1550, \ion{N}{iv} 1720) but they are mainly formed in the wind part of the atmosphere where numerous effects (X-rays, clumping) render their modelling difficult. Photospheric lines are found in the optical range, but, here again, only a few lines are available.

\citet{rgonz11a} have revisited the formation mechanism of \ion{N}{iii} 4634--4640 and have used several \ion{N}{ii}, \ion{N}{iii} and \ion{N}{IV} lines to derive abundances of O stars in the Magellanic Clouds. \citet{martins12} studied the effect of magnetism on the surface abundance of O stars, and based their nitrogen abundance measurements on \ion{N}{ii}, \ion{N}{iii} and \ion{N}{iv} optical lines. Nitrogen is a crucial indicator of chemical enrichment. But more precise constraints can be brought to stellar evolution if the ratio N/C can be derived because N/C is less sensitive to the initial metal content than N/H. For instance the ratio of N/C in the Galaxy to N/C in the SMC is less that 3, while N/H in the Galaxy is 13 times larger than N/H in the SMC. Thus N/C, especially when combined to the N/O ratio \citep[e.g.][]{przy10}, provides stronger constraints on the enrichment and mixing history of CNO material at a given metallicity. 

Unfortunately, carbon lines are even less numerous than nitrogen lines in O stars spectra. In the optical range, the strongest lines are  \ion{C}{iii} 4647--50--51 (hereafter \ion{C}{iii} 4650), \ciii, and \ion{C}{iv} 5801--12. \ion{C}{iii} 4650 is observed in absorption in most O stars. It shows emission only in the earliest O supergiants. In early O dwarfs, it is almost absent. A peculiar class of objects -- the Ofc stars \citep{walborn10} -- displays always \ion{C}{iii} 4650 emission. Stars of the Ofc class encompass Of?p stars which, among other characteristics, have a strong but varying \ion{C}{iii} 4650 emission. Many of them are magnetic \citep{donati06,hubrig08,martins10,wade11}. Ofc stars are found mainly at spectral type O5 but at all luminosity classes \citep{walborn10}. So far, no explanation has been given for the appearance of such \ion{C}{iii} 4650 emission. Its analysis is rendered difficult by the presence of the neighbouring strong \ion{N}{iii} 4634--4640 lines. \ion{C}{iv} 5801--12 is present in most O stars, but is often not observed. In late-type stars it is in absorption, while in early type stars it can exhibit a complex profile with broad emission wings and absorption \citep{conti74,jc12}.

\ciii\ is present in most O stars except the earliest ones \citep{walborn80}. It is in emission in giants and supergiants, and in most dwarfs. The emission is on average stronger in supergiants than in dwarfs \citep{conti74}. Contrary to \ion{C}{iii} 4650, \ciii\ is located in a spectral region free of other lines and hence its profile is well defined. Spectroscopic analysis have generally not considered this line for carbon abundance determination -- atmosphere models often have difficulties reproducing the observed profiles, even if the H, \ion{He}{i}, \ion{He}{ii} and UV CNO lines are well fitted. This indicates that the formation of that line is complex and probably depends on missing or improperly accounted for physics. \citet{nu71} found that the \ciii\ emission was due to a depopulation of the lower level by \ion{C}{iii} 884. Hence, a correct description of the radiative transfer close to 884 \AA\ is required. Since the models used by \citet{nu71} did not include a proper line--blanketing and non--LTE treatment, we revisit the formation of \ciii. We also investigate the formation of \ion{C}{iii} 4650. 

In Section \ref{s_mod} we present the method and models we have used. Sect.\ \ref{form_sing} describes the formation mechanism of \ciii\ and Sect.\ \ref{form_trip} focuses on \ion{C}{iii} 4650. The wind effects are gathered in Sect.\ \ref{s_wind}. We summarize our results in Sect.\ \ref{s_conc}.

%%%%%%%%%%%%%%%%%%%%%%%%%%%%%%%%%%%%%%%%%%%%%%%%%%%%%%%%%%%%%%%%%%%%%%%%%%%%%%%%%%%%%%%%%%%%%%%%%%%%%%%%%%%%%%%%%%%%%%%%%%%%%%%
%%%%%%%%%%%%%%%%%%%%%%%%%%%%%%%%%%%%%%%%%%%%%%%%%%%%%%%%%%%%%%%%%%%%%%%%%%%%%%%%%%%%%%%%%%%%%%%%%%%%%%%%%%%%%%%%%%%%%%%%%%%%%%%
\section{Model description}
\label{s_mod}

\begin{figure*}[]
     \centering
\subfigure[]{ \includegraphics[width=.4\textwidth]{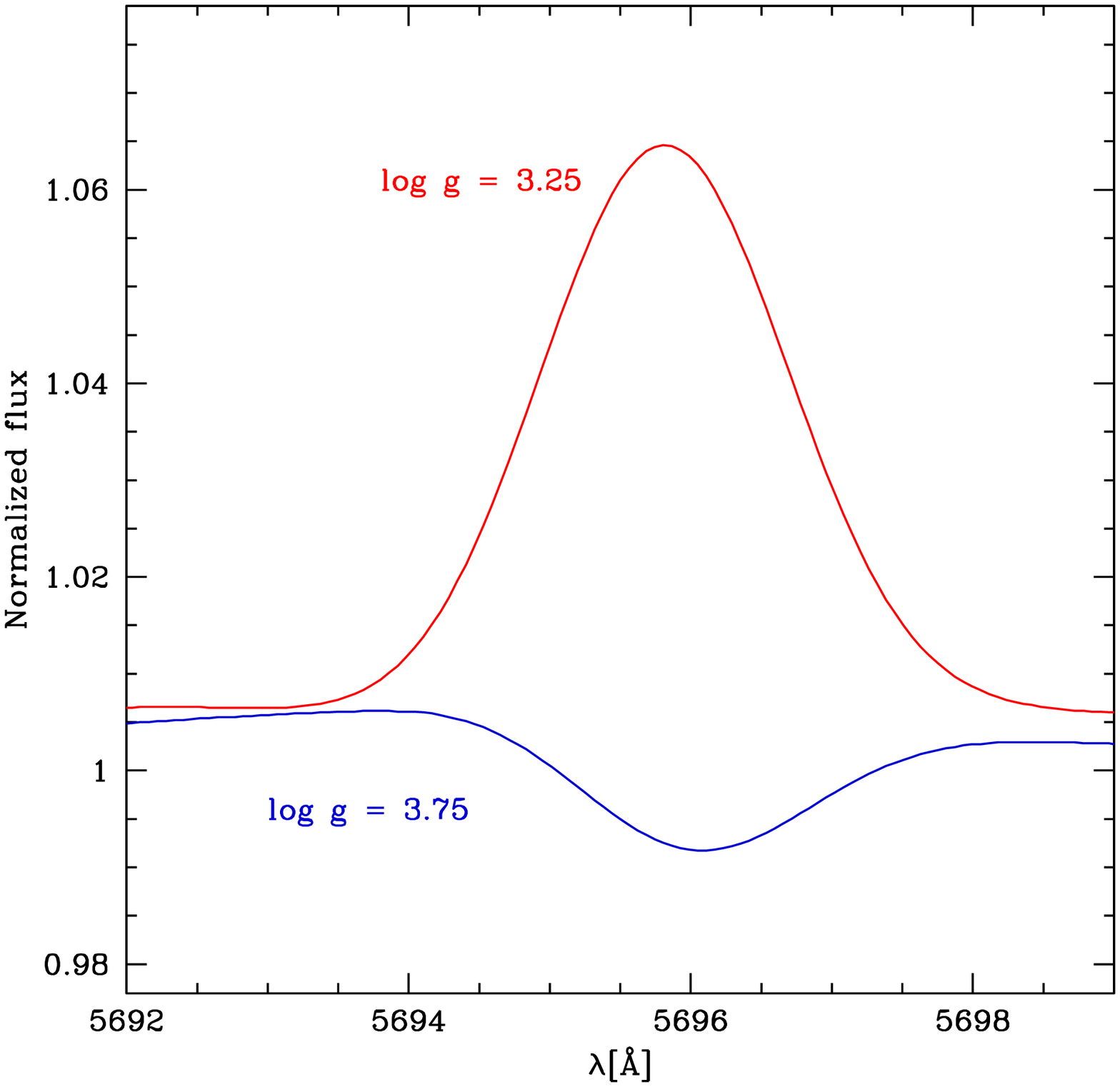}}
     \hspace{0.1cm}
     \subfigure[]{
           \includegraphics[width=.4\textwidth]{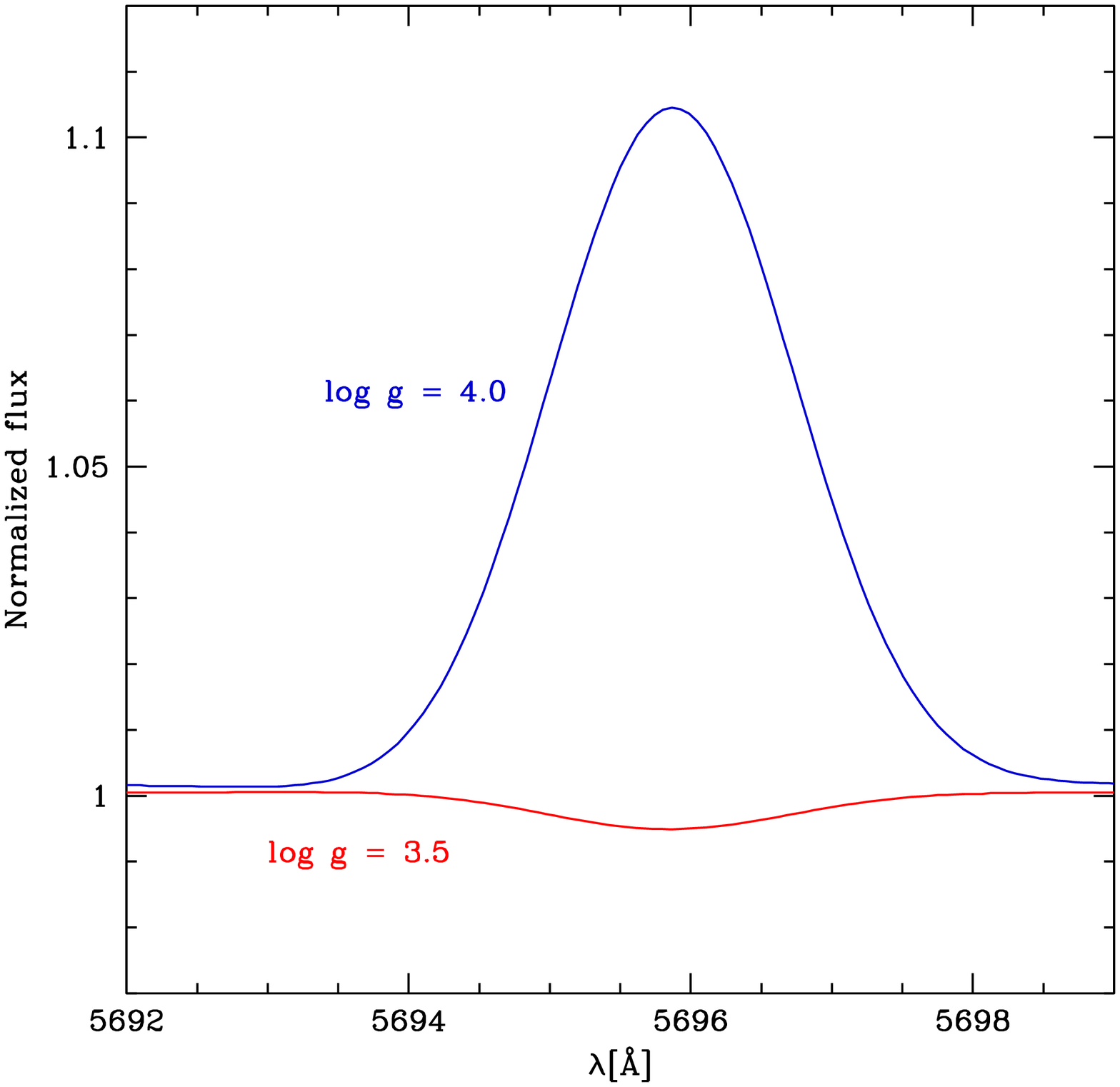}}
     \caption{Spectra of \ciii\ illustrating the dramatic
     dependence of the line profile on surface gravity and effective temperature. The left panel is for \teff=30500\,K while the right panel is for \teff=42000\,K. Particularly striking is the strong dependance of the profile on \logg, and that this dependance is of opposite sign in models of different effective temperature.}
     \label{eff_logg_prof}
\end{figure*}

Our investigation of the formation mechanisms of \ion{C}{iii} 4650 and \ciii\ relied on the computation of model atmosphere and the analysis of the processes populating the levels involved in the \ion{C}{iii} transitions. 
We have used the code CMFGEN \citep{hm98} to compute atmosphere models. CMFGEN produces non-LTE models including winds and line--blanketing. The radiative transfer is solved in the comoving frame and the geometry is spherical. The rate equation and the radiative transfer equation are solved simultaneously and iteratively until convergence of the level populations. A super--level formalism is used to reduce the number of levels involved in the computations (to reduce the amount of memory required). Table \ref{tab_elmts} provides the number of levels used in our computations. Our models include H, He, C, N, O, Ne, Si, S, Ar, Ca, Fe, Ni. The reference solar abundances of \citet{gre07} were adopted. The radiative equilibrium condition sets the temperature structure and the density structure in the inner atmosphere is obtained by iteration. Once the level populations are (roughly) converged, the radiative force is computed and the momentum equation is solved to yield the density structure; the radiative field and level populations are then re-computed until full convergence of the model is achieved. A couple of these hydrodynamical iterations are performed. In the outer atmosphere, the velocity law is given by a classical $\beta$ law \citep[e.g.][]{kud89}. The density and velocity distributions are related via the mass conservation equation. 

The emergent spectrum is computed using CMF\_FLUX \citep{Busch05}, and results from a formal solution of the radiative transfer equation using the atmosphere model provided by CMFGEN. A depth dependent microturbulent velocity, varying from 10 \kms\ in the photosphere to 200 \kms\ in the upper wind, is used. Proper pressure and Stark broadening is included for H, \ion{He}{i}, \ion{He}{ii} and CNO resonance lines. Other lines are treated by Gaussian profiles. 

For our purpose, we selected two sets of model: one with \teff\ = 30500 K and one with 42000 K. They correspond to late and early O stars respectively \citep{msh05}. The atomic data for carbon were taken from various sources: the photoionization rates are from the opacity project \citep{Sea87_OP}\footnote{Data for the opacity project can be obtained from http://cdsweb.u-strasbg.fr/topbase/testop/TheOP.html}, oscillator strengths and the low-temperature dielectronic transition data were kindly computed by Peter Storey (private communication).  For collisional excitation we write the collisional excitation rate (in cgs units) as
$$C(i,k)=8.63 \times 10^{-8} N_e { \Omega(i,k) \over g_i} \sqrt{T \over 10000} e^{-E_{ij}} $$
where $N_e$ is the electron density, $g_i$ the statistical weight of the lower level, $E_{ij}$ the excitation energy, T the temperature, and $\Omega(i,k)$ the dimensionless collision strength. The collisional data for the six lowest levels (singlet and triplet states) are from \citet{bbdk85}; for other collisional data we use the approximation of \cite{Reg62_col} who writes the collision strength for a transition between two levels in-terms of the oscillator strength (for optically allowed transitions).

%%%%%%%%%%%%%%%%%%%%%%%%%%%%%%%%%%%%%%%%%%%%%%%%%%%%%%%%%%%%%%%%%%%%%%%%%%%%%%%%%%%%%%%%%%%%%%%%%%%%%%%%%%%%%%%%%%%%%%%%%%%%%%%
%%%%%%%%%%%%%%%%%%%%%%%%%%%%%%%%%%%%%%%%%%%%%%%%%%%%%%%%%%%%%%%%%%%%%%%%%%%%%%%%%%%%%%%%%%%%%%%%%%%%%%%%%%%%%%%%%%%%%%%%%%%%%%%
%%
\section{\ciii}
\label{form_sing}

\ciii\ (\lsl-\lsu) can either be in absorption or emission, and exhibits a complex variation with \teff\ and \logg. To illustrate these dependencies, we show the \ciii\ profile for \teff = 30500\,K and \logg=3.25 and 3.75 in Fig.\ \ref{eff_logg_prof} (left panel). In the low \logg\ model \ciii\ is in emission, while in the high \logg\ model it is in absorption.  
The opposite behavior is observed in a much  hotter model with \teff\ = 42000\,K -- \ciii\ is in absorption for \logg\ = 3.5 but in emission for  \logg\ = 4.0.

As \ciii\ is primarily photospheric a key quantity controlling the line strength (and whether the line is in emission or absorption) is the line source function. 
An increase in the line source function will weaken an absorption line, possibly driving the line into emission. Conversely, a decrease in the line source function will enhance (weaken) the absorption (emission) profile. 
The line source function can be increased by decreasing the population of the lower level, and/or by increasing the population of the upper level. Thus, to understand the origin of \ciii, we need to understand the transitions populating and depopulating its lower and upper level.

\begin{figure}[h]
\centering
\includegraphics[width=9cm]{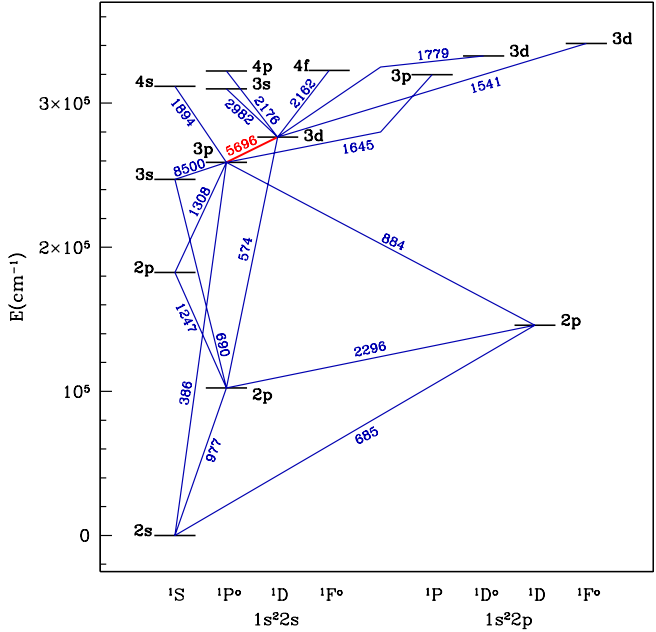}
\caption{Simplified Grotrian diagram for the singlet system of \ion{C}{iii}. The \ciii\ transition is indicated in red. }
\label{gro_singlet}
\end{figure}

 \begin{figure*}[]
     \centering
     \subfigure[]{
%          \label{fig1}
          \includegraphics[width=.45\textwidth]{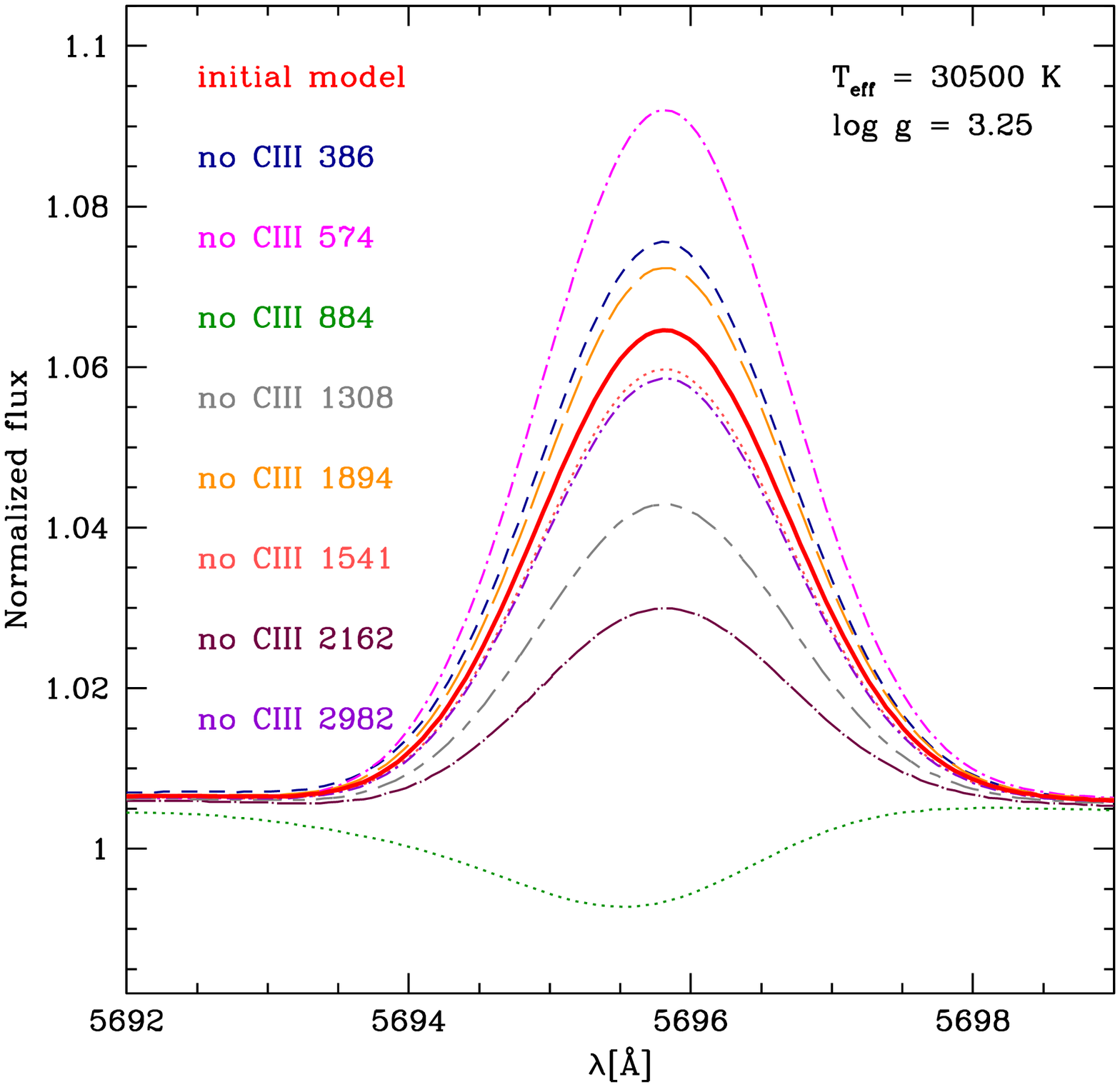}}
     \hspace{0.1cm}
     \subfigure[]{
%          \label{fig2}
          \includegraphics[width=.45\textwidth]{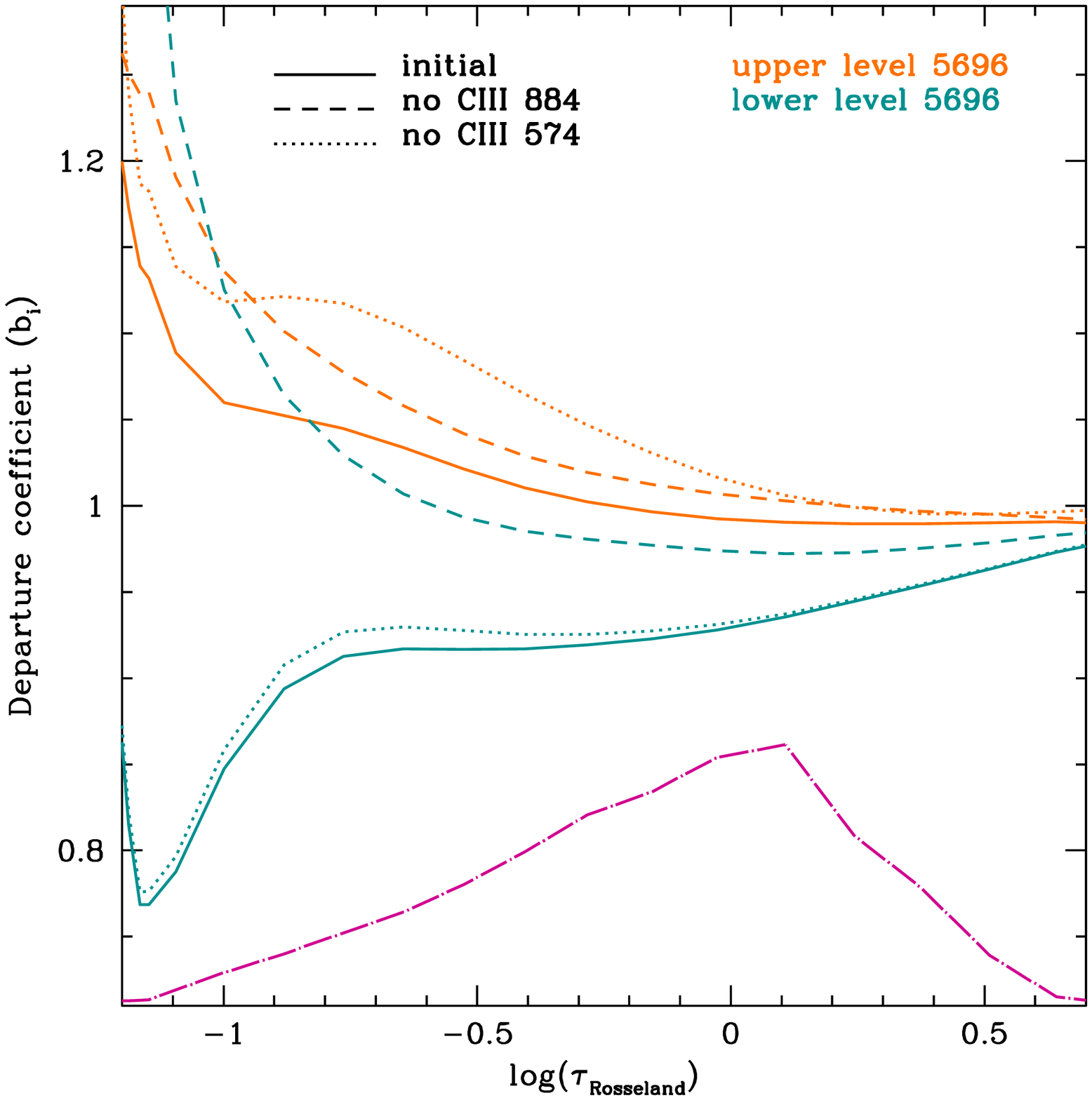}}
     \caption{Effect of the \ion{C}{iii} UV lines on the formation of \ciii. \textit{Left:} Initial \ciii\ spectrum (solid) and spectra from models in which several \ion{C}{iii} UV lines have been artificially removed. \textit{Right:} Departure coefficient of the \ciii\ lower and upper levels in the initial model (solid lines), the model without \ion{C}{iii} 884 \AA\ (dashed lines) and the model without \ion{C}{iii} 574 \AA\ (dotted lines). Departure coefficients are plotted as a function of Rosseland opacity. The dot--dashed line at the bottom indicates the \ciii\ formation region expressed in terms of the contribution function as a function of $\tau_{\rm Rosseland}$.}
     \label{eff_884}
\end{figure*}

Fig.~\ref{gro_singlet} presents a simplified Grotrian diagram of the singlet system of \ion{C}{iii}. We see that the lower and upper levels of \ciii\ (\lsl\ and \lsu\ respectively) are connected to lower levels by several extreme UV transitions, and to upper levels by far UV/UV transitions. Level \lsl\ is also connected to the ground level by a transition at 386\,\AA.

To illustrate which transitions control the lower and upper level populations of \ciii\ we computed different models (all with \teff=30500\,K, \logg=3.25, \mdot\ = $2.0 \times\ 10^{-6}$ \msun\ and \vinf\ = 2000 \kms) in which we selectively removed specific \ion{C}{iii} transitions by reducing their oscillator strength by a factor 10000.  The results are displayed in Fig.\ \ref{eff_884} (left panel). As an example, the dotted green line is the model in which \ion{C}{iii} 884 was removed. This figure highlights the key role of \ion{C}{iii} 884. {\it Its effect is dramatic: \ciii\ switches from emission to absorption when it is removed from the model.} Several other \ion{C}{iii} lines (e.g., \ion{C}{iii} 386 and \ion{C}{iii} 574) also have a significant impact on \ciii, although with a reduced effect compared with \ion{C}{iii} 884.

\begin{figure}[]
\centering
\includegraphics[width=9cm]{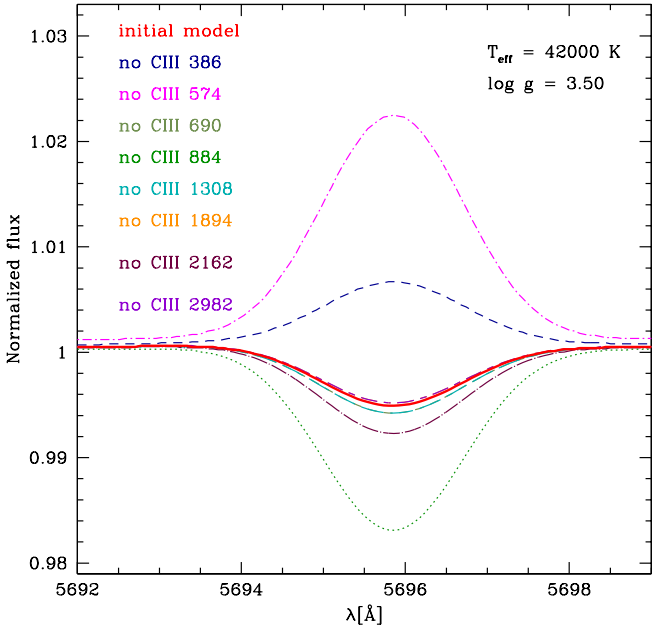}
\caption{Same as left panel of Fig.\ \ref{eff_884} for the model with \teff\ = 42000 K.}
\label{eff_ciiilines_t42}
\end{figure}

\begin{figure*}[]
     \centering
     \subfigure[]{
%          \label{fig2}
          \includegraphics[width=.45\textwidth]{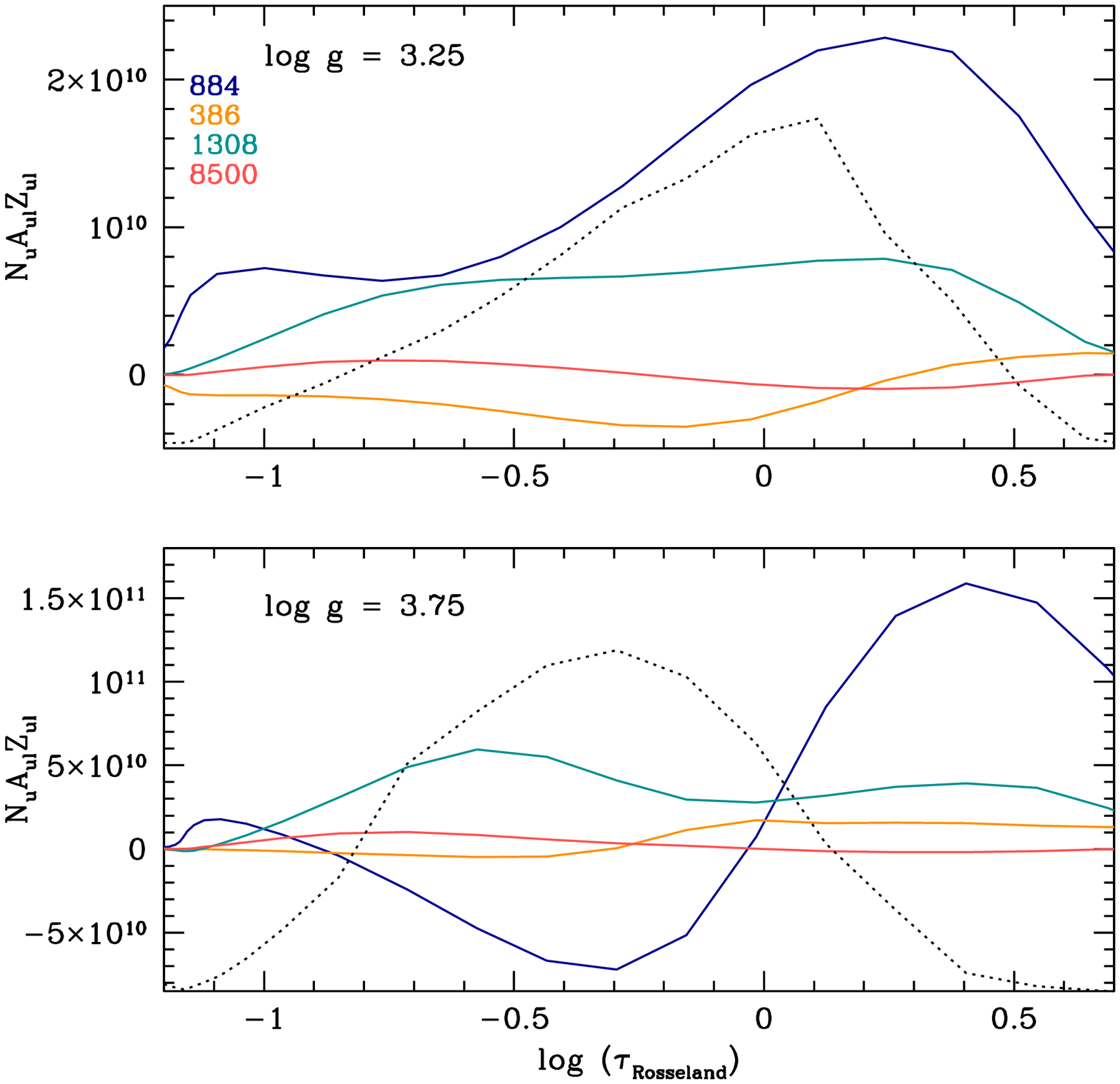}}
     \hspace{0.1cm}
     \subfigure[]{
%          \label{fig3}
          \includegraphics[width=.45\textwidth]{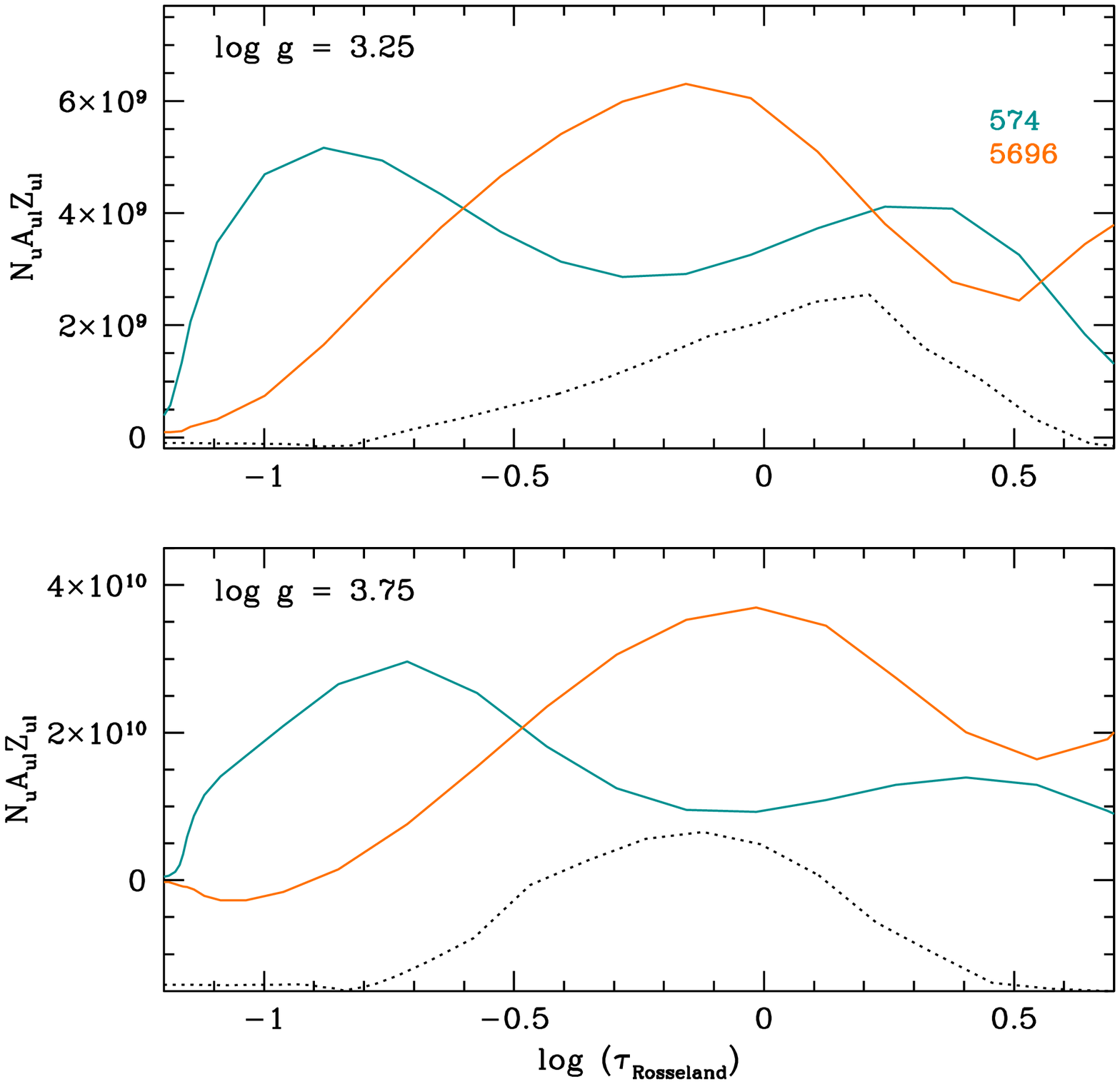}}
     \caption{Effect of \logg\ on the formation of \ciii\ for \teff\ = 30500 K. \textit{Left:} Total radiative rates out of level \lsl\ at \logg\ = 3.25 (upper panel) and \logg\ = 3.75 (lower panel) as a function of Rosseland opacity. Notice how the largest rate is for $\lambda$ 884, and that this rate shows a qualitatively different behavior in the
     two models. The dotted lines indicate the \ciii\ line formation region. \textit{Right}: Total radiative rates out of level \lsu\ at \logg\ = 3.25 (upper panel) and \logg\ = 3.75 (lower panel).}
     \label{eff_logg}
\end{figure*}

The right panel of Fig.\ \ref{eff_884} shows the departure coefficients of levels \lsu\ and \lsl\ (upper and lower levels of \ciii) in the initial model and the models without \ion{C}{iii} 884 and \ion{C}{iii} 574 (i.e., the lines leading to the strongest changes). We see that in the absence of the \ion{C}{iii} 884 \AA\ transition, the departure coefficients are close to 1.0, while in the initial model, the lower level is depopulated compared to the upper level, leading to emission. The total radiative rate\footnote{The total radiative rate is the difference between the number of radiative transitions from the upper level towards the lower level and the number of radiative transitions from the lower level towards the upper level. It is positive when there is a net decay of electrons from the upper state and negative when there is a net excitation of electrons into the upper state.} in \ion{C}{iii} 884 is positive in the initial model (see left panel of Fig.\ \ref{eff_logg}). Thus, \ion{C}{iii} 884 acts as a drain for level \lsl. When we remove it, the drain is suppressed and the \lsl\ population is higher, as seen in Fig.\ \ref{eff_884} (right panel).

Similarly, the total radiative rate in \ion{C}{iii} 574 is also positive (see right panel of Fig.\ \ref{eff_logg}), draining the population of \lsu. When \ion{C}{iii} 574 is removed, \lsu\ is less depopulated, its departure coefficient is higher than in the initial model (Fig.\ \ref{eff_884}) and  \ciii\ emission is strengthened. 

In the model with \teff=42000\,K\footnote{At \teff=42000\,K we have adopted the same mass loss rate as in the low \teff\ model for a better comparison.} the physical conditions are different and the reaction of \ciii\ to the removal of UV lines is slightly different. Fig.\ \ref{eff_ciiilines_t42} shows the effect of different \ion{C}{iii} UV lines on \ciii\ for the model with \teff\ = 42000 K and \logg\ = 3.5. Compared to the left panel of Fig.\ \ref{eff_884} (\teff\ = 30500 K, \logg\ = 3.25), lines have qualitatively similar effects. Removing \ion{C}{iii} 574 leads to a stronger emission than the initial model; removing \ion{C}{iii} 386 also leads to a stronger emission, but not as strong as in the case of \ion{C}{iii} 574; and removing \ion{C}{iii} 884 leads to more absorption. But at 42000 K, it is the removal of \ion{C}{iii} 386 and \ion{C}{iii} 574 (and not \ion{C}{iii} 884) which leads to the switch between absorption and emission. Thus depending on the temperature the \ciii\ emission is driven by different \ion{C}{iii} UV lines. In appendix \ref{s_8500} we present the effect of these lines on the appearance of \ion{C}{iii} 8500 (its upper level is the lower level of \ciii) relevant for future \textit{Gaia} observations.

%%%%%%%%%%%%%%%%%%%%%%%%%%%%%%%%%%%%%%%%%%%%%%%
\subsection{Explanation of \logg\ effect}
\label{form_5696_logg}

Earlier we saw that the \ciii\ profile varies dramatically with both \teff\ and \logg. We can now use our understanding of the key role of \ion{C}{iii} 386, \ion{C}{iii} 574 and \ion{C}{iii} 884 in controlling the strength of \ciii\ to explain this behavior.

\begin{figure}[]
\centering
\includegraphics[width=9cm]{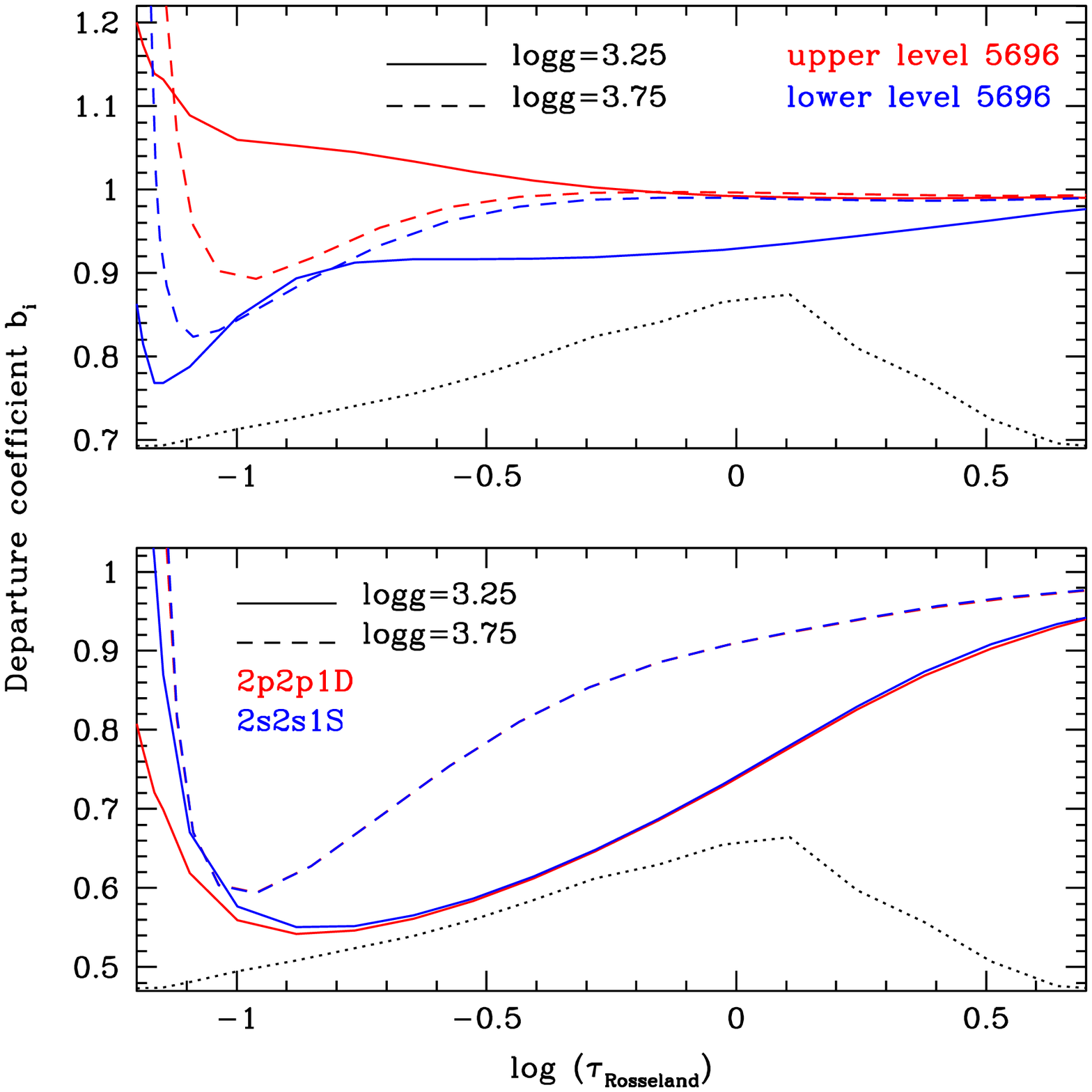}
\caption{Effect of \logg\ on departure coefficients. Departure coefficients are plotted as a function of Rosseland opacity for the levels on \ciii\ (upper panel) and for the fundamental level (2s$^2$~$^1$S) and the lower level of \ion{C}{iii} 884 (2p$^2$~$^1$D). The dotted line shows the \ciii\ line formation region.}
\label{bi_logg}
\end{figure}

\begin{figure*}[]
     \centering
     \subfigure[]{
%          \label{fig2}
          \includegraphics[width=.45\textwidth]{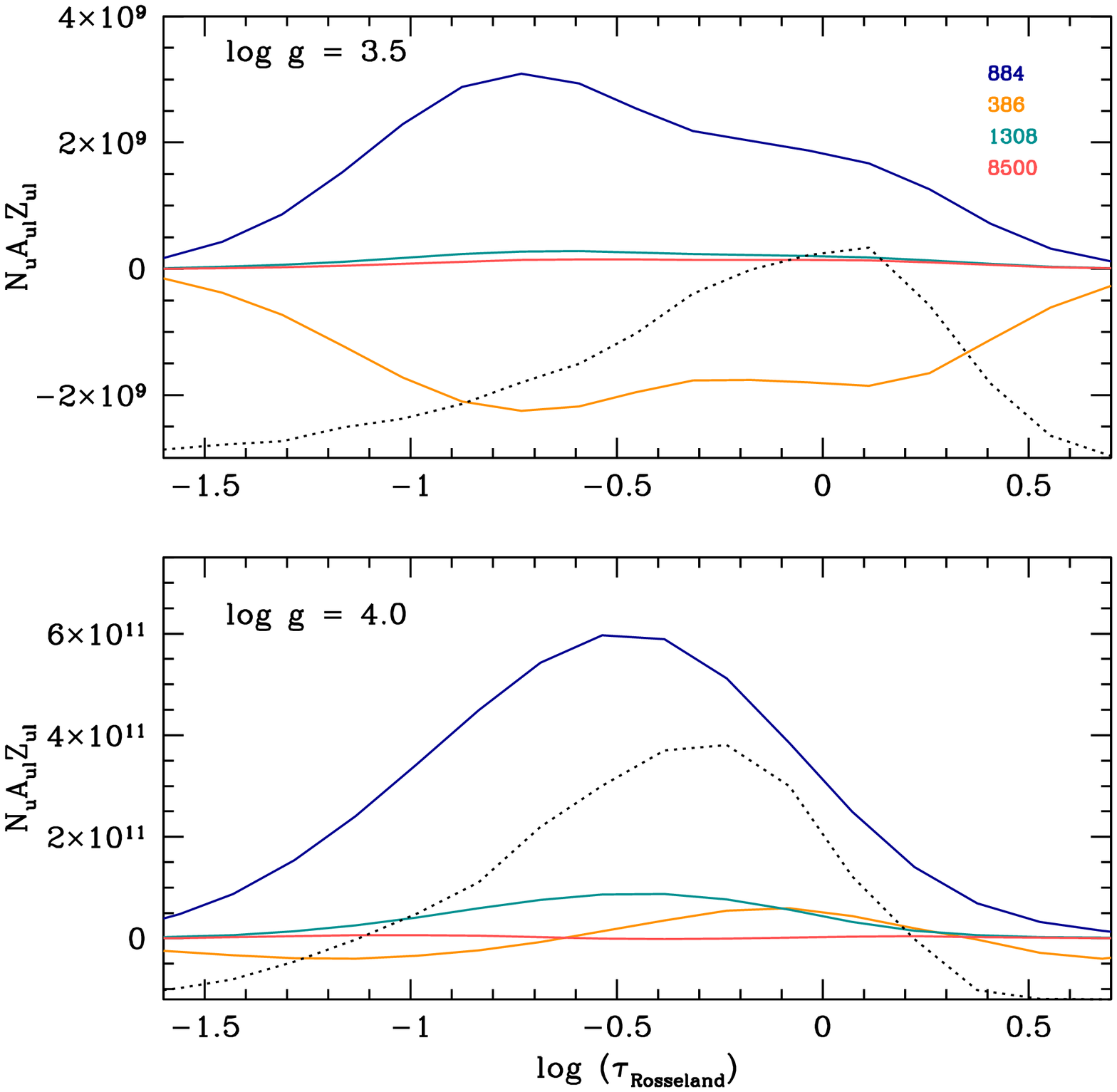}}
     \subfigure[]{
%          \label{fig3}
          \includegraphics[width=.45\textwidth]{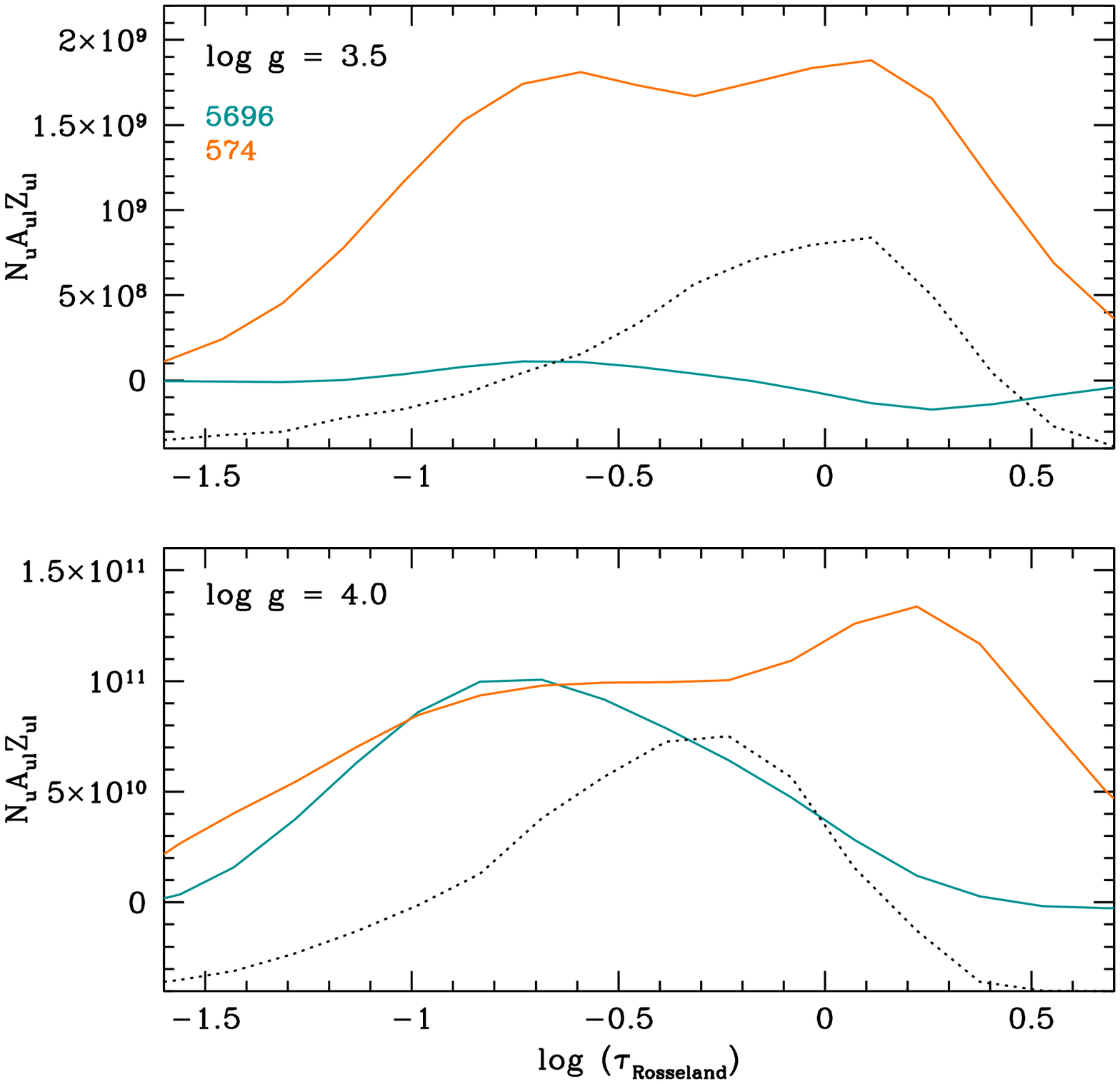}}
     \caption{Same as Fig.\ \ref{eff_logg} but for a model with \teff\ = 42000 K and \logg\ = 3.5 and 4.0.}
     \label{eff_logg_t42}
\end{figure*}

In the left panel of Fig.\ \ref{eff_logg} we show the total radiative rates from level \lsl\ (i.e. the lower lower level of \ciii) for \logg\ = 3.25 and 3.75 at 30500 K. Focusing on the low gravity case, we see that over the \ciii\ formation region the highest rate is observed for the \ion{C}{iii} 884 \AA\ transition. In addition, this rate is positive, indicating a depopulation of the \lsl\ level as seen previously. Turning to the \logg\ = 3.75 model, we see that the \ion{C}{iii} 884 \AA\ radiative rate has a qualitatively different behavior -- in most of the \ciii\ formation region it is negative, and hence it does not act as a significant drain on the population of the lower level of \ciii. Consequently, \ciii\ switches to absorption.

We can understand the differences as follows. A decrease of \logg\ translates into a smaller Lyman jump and to a higher ionizing flux \citep[e.g.][]{ah85}. This also follows from density arguments --- at a given \teff, a low \logg\ model is less dense, hence there are less recombinations and the atmosphere is more highly ionized than a higher \logg\ model. Consequently, the ionization is stronger in the \logg\ = 3.25 model atmosphere. \ion{C}{iv} is the most abundant ion while at \logg\ = 3.75, it is \ion{C}{iii}. In addition, \lsl\ and 2p$^2$~$^1$D (the lower level of \ion{C}{iii} 884) are collisionally coupled.

The switch from absorption to emission for \ciii\ when \logg\ decreases is due to the following chain of events: the ground state and 2p2p~$^1$D are depopulated due to the higher ionization and consequently the downward radiative rate in \ion{C}{iii} 884 has increased; the \ion{C}{iii} 884 drain on the \lsl\ (lower level of \ciii) population increases, which leads to emission in \ciii. Fig.\ \ref{bi_logg} confirms that when \logg\ decreases, the ground state and 2p2p~$^1$D are depopulated, leading to a strong depopulation of \lsl\ through \ion{C}{iii} 884.

At \teff\ = 42000 K, the effect of \logg\ on \ciii\ is opposite (Fig.\ \ref{eff_logg_prof}). The total radiative rates of the low and high \logg\ models at 42000 K are shown in Fig.\ \ref{eff_logg_t42}. \ion{C}{iii} 884 always acts as drain, and drives \ciii\ into emission. However, the resonance line \ion{C}{iii} 386 switches from near radiative equilibrium at \logg\ = 4.0 to a net negative rate at \logg\ = 3.5 (left panel of Fig.\ \ref{eff_logg_t42}), implying that it populates the \lsl\ level. This compensates for the \ion{C}{iii} 884 drain, and \ciii\ is in weak absorption. 

In conclusion, the reaction of \ciii\ to changes in \logg\ is directly related to the line formation mechanism and the key role of the \ion{C}{iii} UV lines. As the temperature increases, the peak of the spectral energy distrubution shifts to shorter wavelengths and consequently higher energy \ion{C}{iii} UV lines become more and more important at the expense of lower energy transitions.

%%%%%%%%%%%%%%%%%%%%%%%%%%%%
\subsection{Dielectronic recombinations}

\begin{figure}[h]
\centering
\includegraphics[width=9cm]{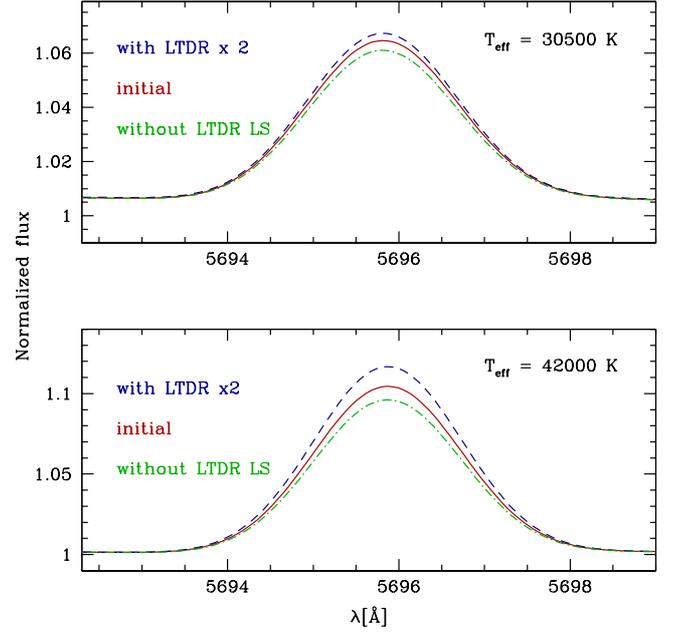}
\caption{Effect of low temperature dielectronic recombination (LTDR) on \ciii\ in the models with \teff\ = 30500 K (upper panel) and 42000 K (lower panel). The red solid curve shows the standard model profile, while the blue dashed profile shows the effect of
``double counting" the influence of LTDR. The green dot--dashed line is the standard model in which the LTDR from levels forbidden to autoionize in LS coupling have been omitted.}
\label{eff_dieTT}
\end{figure}

So far we have focused on detailed non--LTE radiative transfer effects to explain the morphology of \ciii. Dielectronic recombination is another mechanism which might have an influence on the appearance of \ciii. As low temperature dielectronic recombination (LTDR) for levels permitted to autoionize in LS-coupling is handled via resonances in the photoionization cross-sections they cannot be easily omitted from the calculations. Instead, we run a model in which we utilize the Einstein $A$ values computed using intermediate coupling by Peter Storey (private communication) to add important LTDR transitions into the photoionization cross-sections. Although this means that we are exaggerating their influence (since we are double counting) we can still use the change in line profile to infer whether LTDR is important. Fig.\ \ref{eff_dieTT} shows the effect of low temperature dielectronic recombinations on the appearance of \ciii\ at high and low effective temperature. The additional dielectronic recombinations, when included, lead to a slight increase of the emission line, but quantitatively, the effect is smaller than that of UV lines (see Figs.\ \ref{eff_884} and \ref{eff_ciiilines_t42}). We also run a test model in which we switch off LTDR for transitions not permitted to autoionize in LS coupling. These were also found to have only a small effect on \ciii. Thus LTDR is not a viable mechanism to explain the observed behavior of \ciii\ in O stars. Similar results are found for \ion{N}{iii} 4634--4640 by \citet{rgonz11a}.

%%%%%%%%%%%%%%%%%%%%%%%%%%%%%%%%%%%%%%%%%%%%%%%%%%%%%%%%%%%%%%%%%%%%%%%%%%%%%%%%%%%%%%%%%%%%%%%%
\subsection{Line blanketing effects}

%%%%%%%%%%%%%%%%%%%%%%%%
\subsubsection{Microturbulence and metallic line interactions}
\label{LB_5696}

\begin{figure}[]
\centering
\includegraphics[width=9cm]{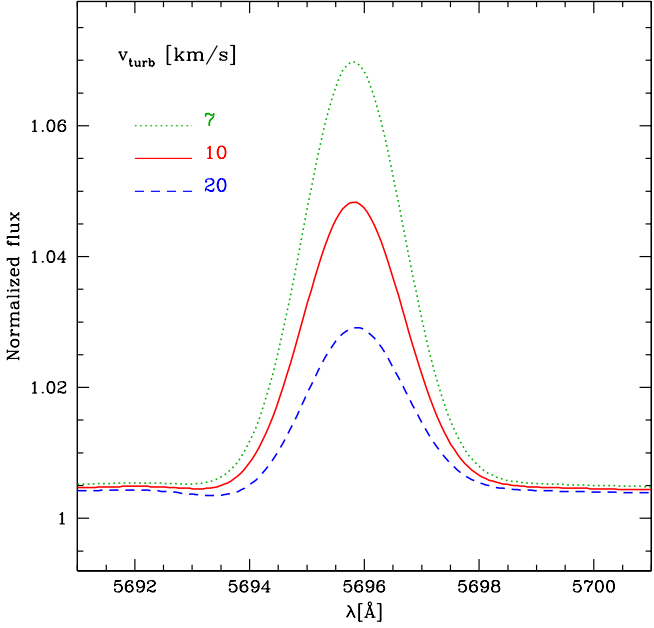}
\caption{Effect of the microturbulent velocity on the \ciii\ emission strength in three CMFGEN models. The microturbulent velocity has been changed only in the atmosphere computation. The spectra have all been computed with \vturb\ increasing from 10 to 200 \kms\ throughout the wind. }
\label{spec_vturb}
\end{figure}

In Fig.\ \ref{spec_vturb} we show the effect of a change of the microturbulent velocity in the atmosphere model (and not in the calculation of the spectrum) on the \ciii\ line. A reduction of \vturb\ translates into a \textit{stronger} emission. Modifying \vturb\ affects the efficiency of line--blocking and thus the temperature structure and EUV radiation field. But different microturbulent velocities also imply different metallic line width and thus different line overlap, which can affect radiative transfer in UV \ion{C}{iii} lines.

We have seen in section \ref{form_sing} that the \ion{C}{iii} 386, \ion{C}{iii} 574 and \ion{C}{iii} 884 lines significantly influence the strength of \ciii. Depending on the microturbulent velocity the lines have a typical width between 15 and 30 \kms, and hence will interact with other lines whose separations are less than 15--30\,\kms. To test the effect of metallic lines on the \ion{C}{iii} UV features and thus on \ciii, we have run test models with  \teff\ = 30500\,K. We selectively removed metallic features (\ion{Fe}{vi} 386.186, \ion{Fe}{iv} 386.211, \ion{Fe}{v} 386.262, \ion{Fe}{iv} 574.229, \ion{Fe}{vi} 574.230, \ion{Fe}{iv} 574.232, \ion{Ar}{iv} 574.243, \ion{Ni}{iv} 574.291, \ion{N}{iii} 574.298, \ion{N}{iii} 884.466, \ion{Fe}{iv} 884.497, \ion{Fe}{iv} 884.520, \ion{S}{v} 884.530, and \ion{Fe}{v} 884.578) in the model atmosphere and computed the resulting \ciii\ line profile. Atomic data for the transitions found to influence \ciii\ and \ion{C}{iii}\ are discussed in Appendix \ref{ap_met_line}.

\begin{figure}[]
\centering
\includegraphics[width=9cm]{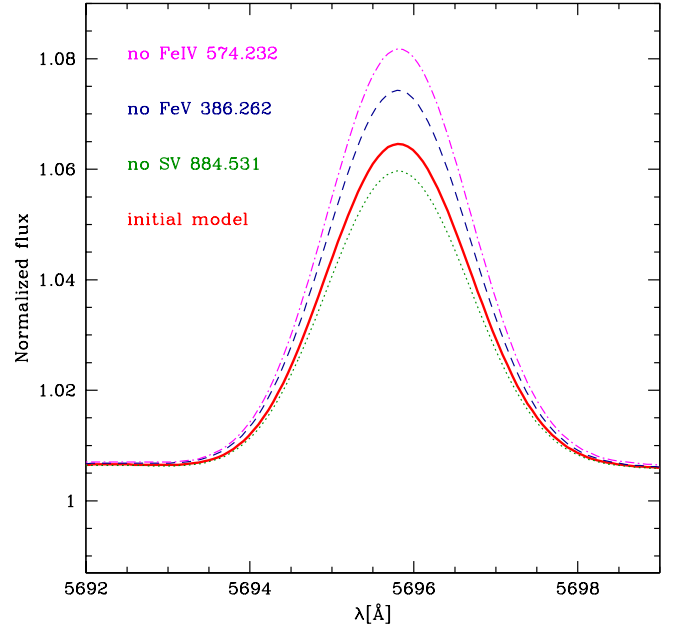}
\caption{Effect of metallic lines on \ciii. The solid line is the initial spectrum. The dashed line is a model in which the oscillator strength of \ion{Fe}{v} 386.262 \AA\ is artificially reduced by a factor 10000. The dot-dashed (dotted) line is the initial model in which \ion{Fe}{iv} 574.232 \AA\ (\ion{S}{V} 884.531 \AA) is reduced by the same factor.}
\label{spec_felines}
\end{figure}

\begin{figure}[]
\centering
\includegraphics[width=9cm]{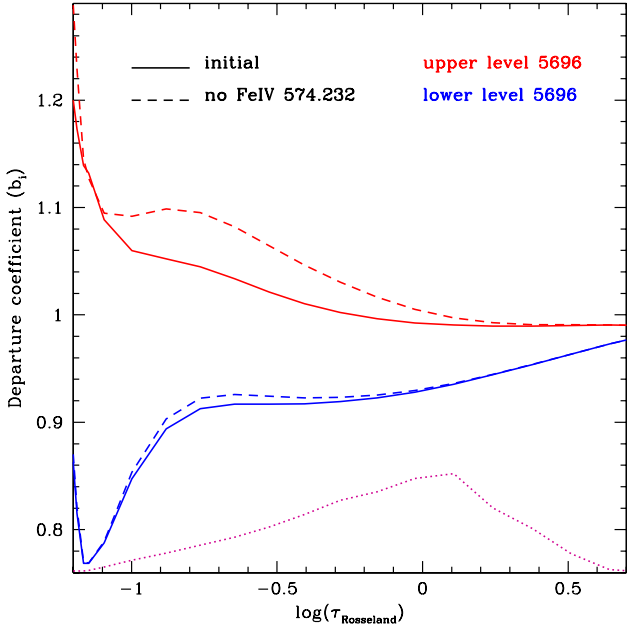}
\caption{Effect of a change in the oscillator strength of \ion{Fe}{iv} 574.232 \AA\ on the departure coefficient (plotted as a function of Rosseland opacity) of the levels of the \ciii\ transition. The red curves correspond to the upper level and the blue ones to the lower level. The solid lines are for the initial model. The dashed lines are the level populations in the model with a reduced \ion{Fe}{iv} 574.232 line. The dotted line is the \ciii\ formation region.}
\label{bi_effect_feiv574}
\end{figure}

\begin{figure}[]
\centering
\includegraphics[width=9cm]{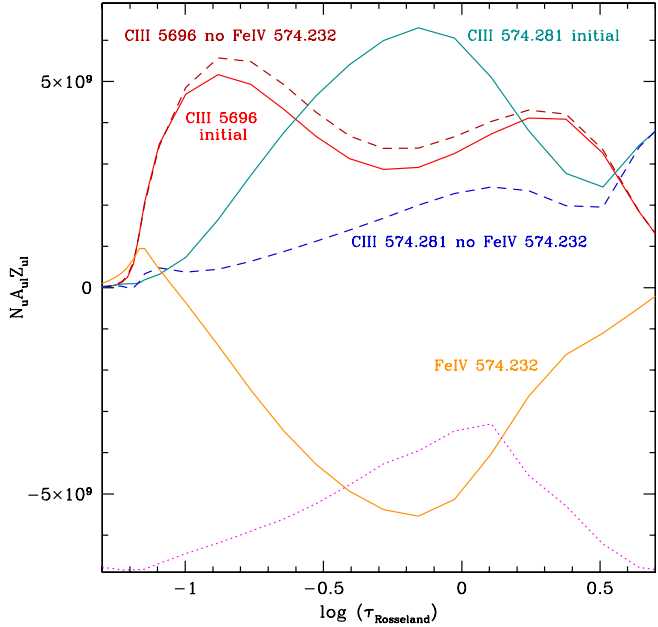}
\caption{Effect of the \ion{Fe}{iv} 574.232 \AA\ line on the total radiative rates of several transitions. The solid lines are for the initial model, the dashed lines for the model in which the iron transition has been removed in the atmosphere calculation. The dotted line indicates the \ciii\ line formation region in the initial model.}
\label{rate_eff_574}
\end{figure}

In all but three cases, which are shown in  Fig.\ \ref{spec_felines}, the \ciii\ feature is basically unchanged. The model with \ion{Fe}{iv} 574.232 removed (dot-dashed) has a stronger \ciii\ than the initial model (solid). The explanation for this emission increase can be inferred from Figs.\ \ref{bi_effect_feiv574} and \ref{rate_eff_574}. In Fig.~\ref{bi_effect_feiv574}, the departure coefficients of the upper and lower levels of \ciii\ are shown for both the initial model and the model without \ion{Fe}{iv} 574.232. Initially, the lower level is more depopulated than the upper level, explaining the emission. When \ion{Fe}{iv} 574.231 is removed, the upper level becomes more populated while the departure coefficient of the lower level is barely changed. This translates into the stronger emission seen in Fig.\ \ref{spec_felines}. 

The reason for the enhanced emission can be understood from Fig.\ \ref{rate_eff_574} which shows the total radiative rates in \ion{C}{iii} 574.281, \ciii\ and \ion{Fe}{iv} 574.231. The total rate in \ion{C}{iii} 574.281 is positive in the initial model while the rate in the iron transition is negative. Consequently the \ion{Fe}{iv} 574.232 transition removes photons from the \ion{C}{iii} 574.281 line. These photons are not available for re-absorption in the \ion{C}{iii} 574.281 and hence the population of \lsu, the upper level of \ciii, is reduced. When the \ion{Fe}{iv} 574.232 line is artificially removed, it does not ``steal'' photons any more from the \ion{C}{iii} 574.281 transition, resulting in a smaller depopulation of \lsu, as seen in Fig.\ \ref{bi_effect_feiv574}, and the emission in \ciii\ is stronger.

The scenario presented above also partly\footnote{The changes in the atmospheric structure -- such as the temperature structure alteration -- also play a role.} explains why a lower microturbulent velocity results in a stronger \ciii\ emission. In the case of a large \vturb, the line overlap between \ion{C}{iii} 574.281 and \ion{Fe}{iv} 574.232 is wider, favoring the drain effect described above. When \vturb\ is reduced, the reduction of the overlap translates in a less efficient drain of the \lsu\ level population, implying a stronger \ciii\ emission.

In Fig.\ \ref{spec_felines}, we have also plotted the spectrum resulting from a model in which \ion{Fe}{v} 386.262 \AA\ was reduced artificially. As for the \ion{Fe}{iv} 574.232 line, the result is a stronger emission. This \ion{Fe}{v} line can interact with \ion{C}{iii} 386 (Fig.\ \ref{gro_singlet}). The total radiative rates in the initial model are negative for both lines (\ion{Fe}{v} 386.262 and \ion{C}{iii} 386). This means that the iron line absorbs photons that could in principle be absorbed by the carbon line. When the iron line is removed, the photons are effectively absorbed by the carbon line. This excess of absorption could in principle overpopulate the \lsl\ level and lead to a weaker emission. But in practice, the \lsl\ population is reduced. This is explained by the need to maintain the ground level population constant, as indicated by our models (the ground level population being controlled by ionization/recombination processes). Since in the absence of \ion{Fe}{v} 386.262 the absorption in \ion{C}{iii} 386 is stronger and since the ground level population is unchanged, an increase of the downward transition is needed to maintain the ground level rate equation equilibrium. Consequently, the \lsl\ is \textit{depopulated}, leading to a stronger emission. Here again, this mechanism explains that a smaller microturbulent velocity leads to less absorption by \ion{Fe}{v} 386.262 and thus to a stronger \ciii\ emission. 

Finally, Fig.\ \ref{spec_felines} also shows the effect of the \ion{S}{V} 884.531 line on \ciii. Removing this feature translates into a reduction of the \ciii\ emission. The explanation comes from the Grotrian diagram in Fig.\ \ref{gro_singlet} and the effect of the \ion{C}{iii} 884 transition on \ciii\ (at \teff\ = 30500 K). The total radiative rate in the \ion{S}{V} line is negative throughout the \ciii\ line formation region, while the \ion{C}{iii} 884 \AA\ line rate is positive. Thus the sulfur line enhances the drain of \lsl\ through \ion{C}{iii} 884. By removing the sulfur line, we reduce the drain, thus re-populate the \lsl\ level, and consequently reduce the \ciii\ emission.

%%%%%%%%%%%%%%%%%%%%%%%%
\subsubsection{Metallicity effects}

\begin{figure*}[t]
     \centering
     \subfigure[]{
%          \label{fig1}
          \includegraphics[width=.45\textwidth]{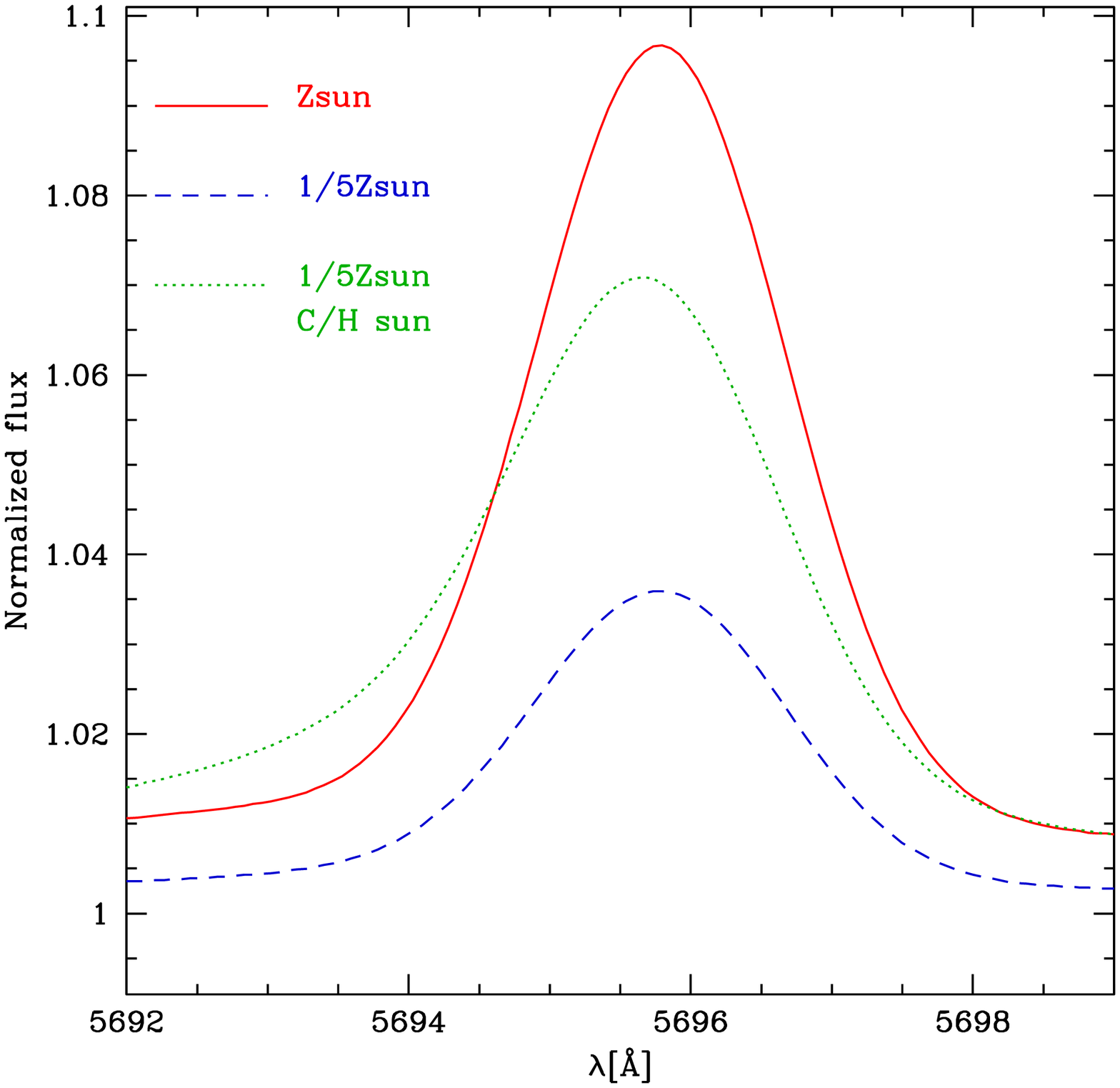}}
     \hspace{0.1cm}
     \subfigure[]{
%          \label{fig2}
          \includegraphics[width=.45\textwidth]{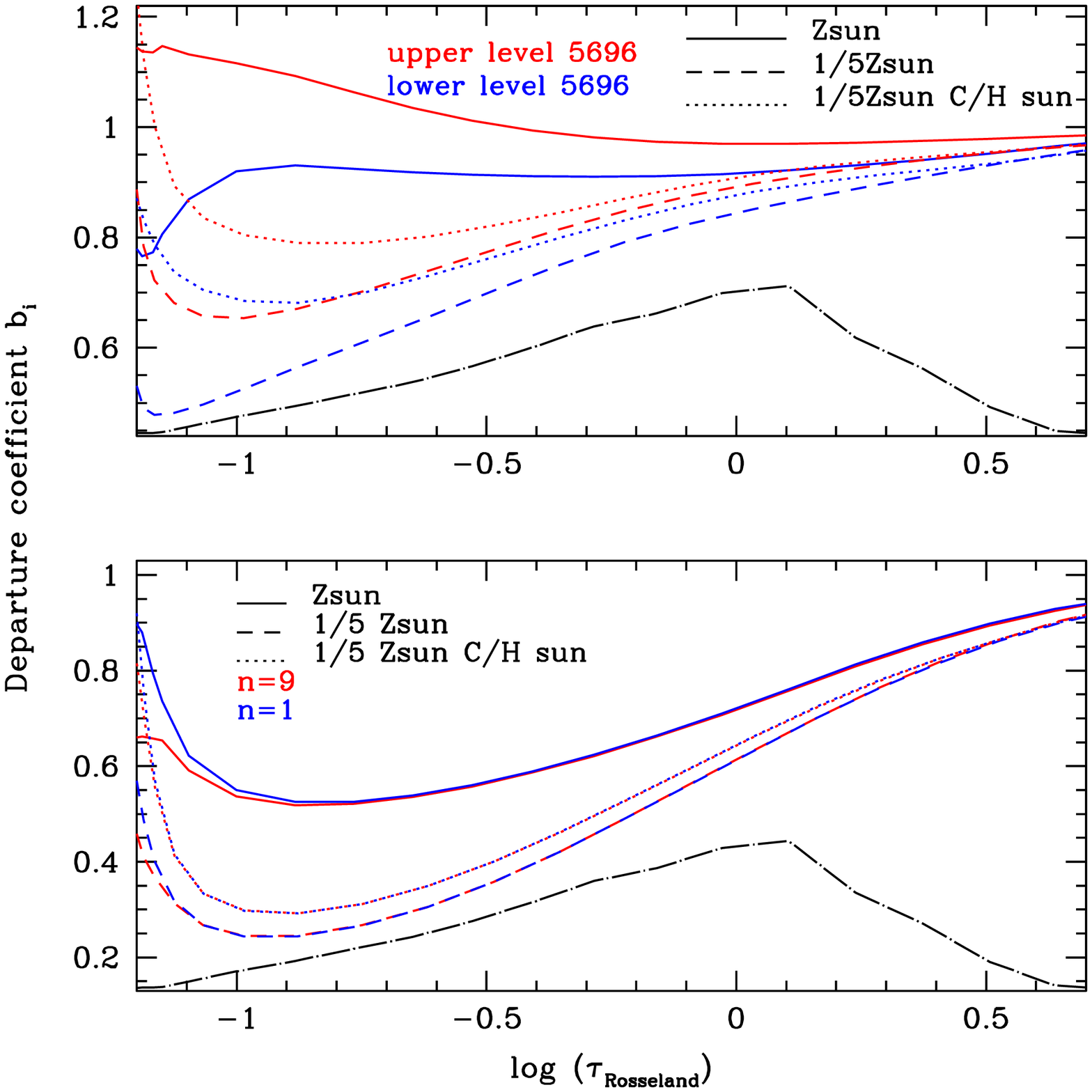}}
     \caption{Effect of the metallicity Z on the formation of \ciii\ at \teff\ = 30500 K. \textit{Left:} Spectra of \ciii\ at Z$_{\odot}$ (solid), 1/5$^{th}$ Z$_{\odot}$ (dashed) and 1/5$^{th}$ Z$_{\odot}$ but with C/H at the solar value (dotted). \textit{Right:} Departure coefficient of the \ciii\ lower and upper levels at  Z$_{\odot}$ (solid lines), 1/5$^{th}$ Z$_{\odot}$ (dashed lines) and 1/5$^{th}$ Z$_{\odot}$  with C/H at the solar value (dotted). The dot--dashed line indicates the line formation region.}
     \label{eff_Z}
\end{figure*}

We have seen in Section \ref{form_5696_logg} that \ciii\ was indirectly sensitive to the ionizing flux. We thus expect that any reduction of the Lyman continuum flux will affect the \ciii\ morphology. Since an increase of the metal content severely reduces the ionizing fluxes \citep[e.g.][]{sdk97}, and thus the ionization structure, we expect metallicity to have a significant impact on the appearance of \ciii.

Fig.\ \ref{eff_Z} shows the influence of the metallicity Z on the \ciii\ line profile and departure coefficients at \teff\ = 30500 K. We have computed models with a global metallicity reduced by a factor five compared to the initial, solar metallicity model. We have also computed a model with a global metallicity of 1/5$^{th}$ the solar metallicity, but with the carbon abundance fixed to the solar value. As expected, at low metallicity the ground state level is depopulated (see right panel of Fig.\ \ref{eff_Z}). The collisional coupling with the 2p2p~$^{1}$D level leads to a strong drain through \ion{C}{iii} 884 \AA\ and to a depopulation of \lsl. But at the same time, we observe a strong depopulation of \lsu\ as well, so that the net outcome of a reduction of Z is a decrease of the \ciii\ emission. This is not just an effect of the scaling of the emission strength with the carbon abundance, since in the model with a solar C/H but reduced metallicity we also observe a reduction of the line intensity. The reason is the depopulation of \lsu. It is due to a stronger radiative rate in the \ion{C}{iii} 574 line at low metallicity. This is explained by the reduced strength of the \ion{Fe}{iv} 574.232 line and consequently its reduced interaction with \ion{C}{iii} 574 as described in Sect.\ \ref{LB_5696}. 

Appendix \ref{ap_Z_t42} shows how \ciii\ reacts to a change of metallicity for \teff\ = 42000 K. A combination of lower ionizing flux and reduced metallic line interaction explains the observed behavior. This highlights the complexity in predicting the shape of \ciii. Metallicity can also affect the line formation by its influence on the wind strength (see Sect.\ \ref{s_wind}).

%%%%%%%%%%%%%%%%%%%%%%%%%%%%%%%%%%%%%
\subsubsection{Influence of  \ion{N}{iv} $\lambda$322}
\citet{gauzit66} argued that \ciii\ emission could be due to the interplay between \ion{N}{iv} 322 and \ion{C}{iii} 322 between levels 2s$^2$~$^1$S and 2p3s\,$^1$P$^{\scriptsize \rm o}$ \footnote{For clarity and given the conclusion of our test, we have not shown these transitions in Fig.\ \ref{gro_singlet}}, the latter being connected to \lsu\ (the upper level of \ciii) by a transition at 2982 \AA\ \citep[see discussion in][]{nu71}. We have tested this possibility by removing \ion{C}{iii} 2982. The result, shown in Fig.\ \ref{eff_884}, indicates that the 2982 \AA\ feature plays little role in populating \lsu. Hence, inducing the \ion{C}{iii} 322 transition and decay through \ion{C}{iii} 2982 by \ion{N}{iv} 322 is not responsible for the appearance of \ciii\ emission.   

\vskip 15pt
In conclusion, we have identified a general mechanism responsible for the morphology of \ciii. It is the influence of UV \ion{C}{iii} transitions on the population of the upper and lower levels of \ciii\ that play a key role. The main lines are \ion{C}{iii} 386, \ion{C}{iii} 574, and \ion{C}{iii} 884. At low \teff, \ion{C}{iii} 884 is crucial for determining the lower level population, while at high \teff, \ion{C}{iii} 386 is more important. Subtle line--blanketing effects modify the strength of the \ciii\ emission. Obtaining an excellent fit to \ciii\ is an extremely challenging task since it requires the knowledge of several stellar parameters including the microturbulent velocity, the metallicity,  as well as accurate wavelengths and oscillator strength of several weak metal lines.

%%%%%%%%%%%%%%%%%%%%%%%%%%%%%%%%%%%%%%%%%%%%%%%%%%%%%%%%%%%%%%%%%%%%%%%%%%%%%%%%%%%%%%%%%%%%%%%%%%%%%%%%%%%%%%%%%%%%%%%%%%%%%%%
%%%%%%%%%%%%%%%%%%%%%%%%%%%%%%%%%%%%%%%%%%%%%%%%%%%%%%%%%%%%%%%%%%%%%%%%%%%%%%%%%%%%%%%%%%%%%%%%%%%%%%%%%%%%%%%%%%%%%%%%%%%%%%%
%%
\section{\ion{C}{iii} 4647--50--51}
\label{form_trip}

We now turn to the study of \ion{C}{iii} 4650. We follow the same method as that used to identify the formation mechanism of \ciii.

%%%%%%%%%%%%%%%%%%%%%%%%%%%%%%%%%%%%%%%%%%%%%%%%%%%%%%%%%%%%%%%%%%%%%%%%%%%%%%%%%%%%%%%%%%%%%%%%%%%%%%
\subsection{Line formation mechanism}
\label{form_4650}

\begin{figure}[h]
\centering
\includegraphics[width=9cm]{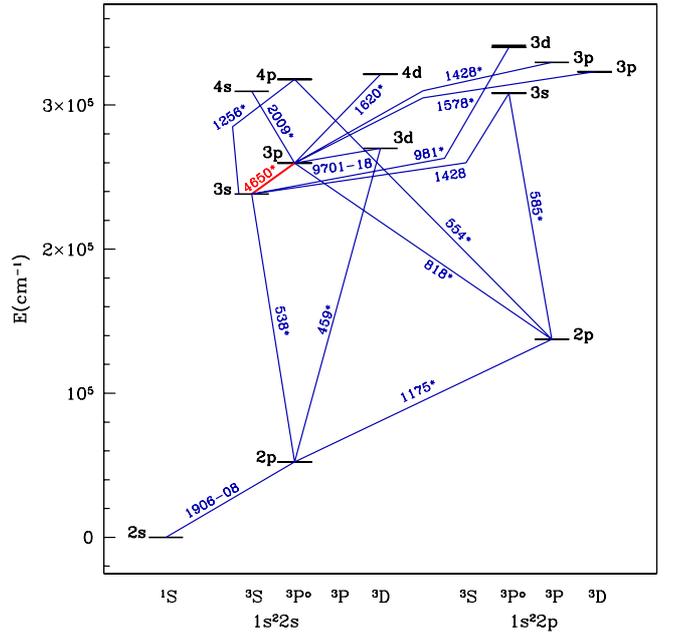}
\caption{Simplified Grotrian diagram for the triplet system of \ion{C}{iii}. The \ion{C}{iii} 4647-50-51 transitions are shown in red. The star symbols indicate multiplets. }
\label{gro_triplet}
\end{figure}

The \ion{C}{iii} 4650 lines are produced by the triplet system of \ion{C}{iii}. They are the triplet equivalent of the \ion{C}{iii} 8500 transition in the singlet system (i.e., a transition between levels 3s and 3p). The triplet equivalent of \ciii\ is \ion{C}{iii} 9710. Fig.\ \ref{gro_triplet} shows the simplified Grotrian diagram for this system.

%%%%%%%%%%%%%%%%%%%%%%%%%%%%%%%%%%%%%
\subsubsection{Effects of UV lines}

\begin{figure}[]
\centering
\includegraphics[width=9cm]{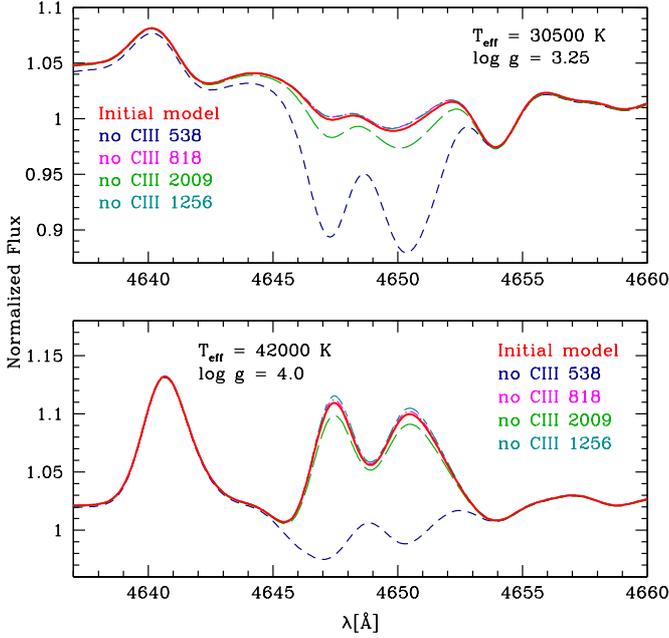}
\caption{Effect of \ion{C}{iii} UV lines on \ion{C}{iii} 4647--50--51 in the model with \teff\ = 30500 K (upper panel) and 42000 K (lower panel). The solid line is the initial model. The other lines are models in which a given \ion{C}{iii} UV line has been removed by reducing its oscillator strength by a factor 10000. }
\label{eff_ciiilines_4650}
\end{figure}

To understand the formation of \ion{C}{iii} 4650, we proceeded as for \ciii, looking for key transitions that control the upper and lower level  populations. Fig.\ \ref{eff_ciiilines_4650} shows the \ion{C}{iii} 4647-50-51 profiles in the low (upper panel) and high (lower panel) \teff\ case. The solid line is the initial model, while the other lines correspond to models in which specific \ion{C}{iii} UV lines connected to either the upper of the lower level of  \ion{C}{iii} 4650 have been removed (by reducing their oscillator strength by a factor 10000). Only the models with the largest changes are shown. The following lines were tested: \ion{C}{iii} 538, \ion{C}{iii} 818, \ion{C}{iii} 981, \ion{C}{iii} 1175, \ion{C}{iii} 1256, \ion{C}{iii} 1428, \ion{C}{iii} 1578, \ion{C}{iii} 1620, \ion{C}{iii} 1908, \ion{C}{iii} 2009.

The \ion{C}{iii} 538 \AA\ transition is found to play a crucial in the formation of the \ion{C}{iii} 4647-50-51 line, since removing it leads to a strong increase of the absorption. The \ion{C}{iii} 2009 line also has a non negligible role on the shape of  \ion{C}{iii} 4650, but its impact is much weaker than that of \ion{C}{iii} 538. The same results are obtained in the model with \teff\ = 42000 K -- all models have \ion{C}{iii} 4650 in emission, except the model in which \ion{C}{iii} 538 is removed and where absorption is observed. From the Grotrian diagram (Fig.\ \ref{gro_triplet}) we can suspect \ion{C}{iii} 538 to be a drain for level \ltl. Inspection of the departure coefficient of \ltl\ in the initial model, and the model without \ion{C}{iii} 538 (not shown), confirm that \ltl\ is more populated when \ion{C}{iii} 538 is absent.

\begin{figure*}[t]
     \centering
     \subfigure[]{
%          \label{fig1}
          \includegraphics[width=.45\textwidth]{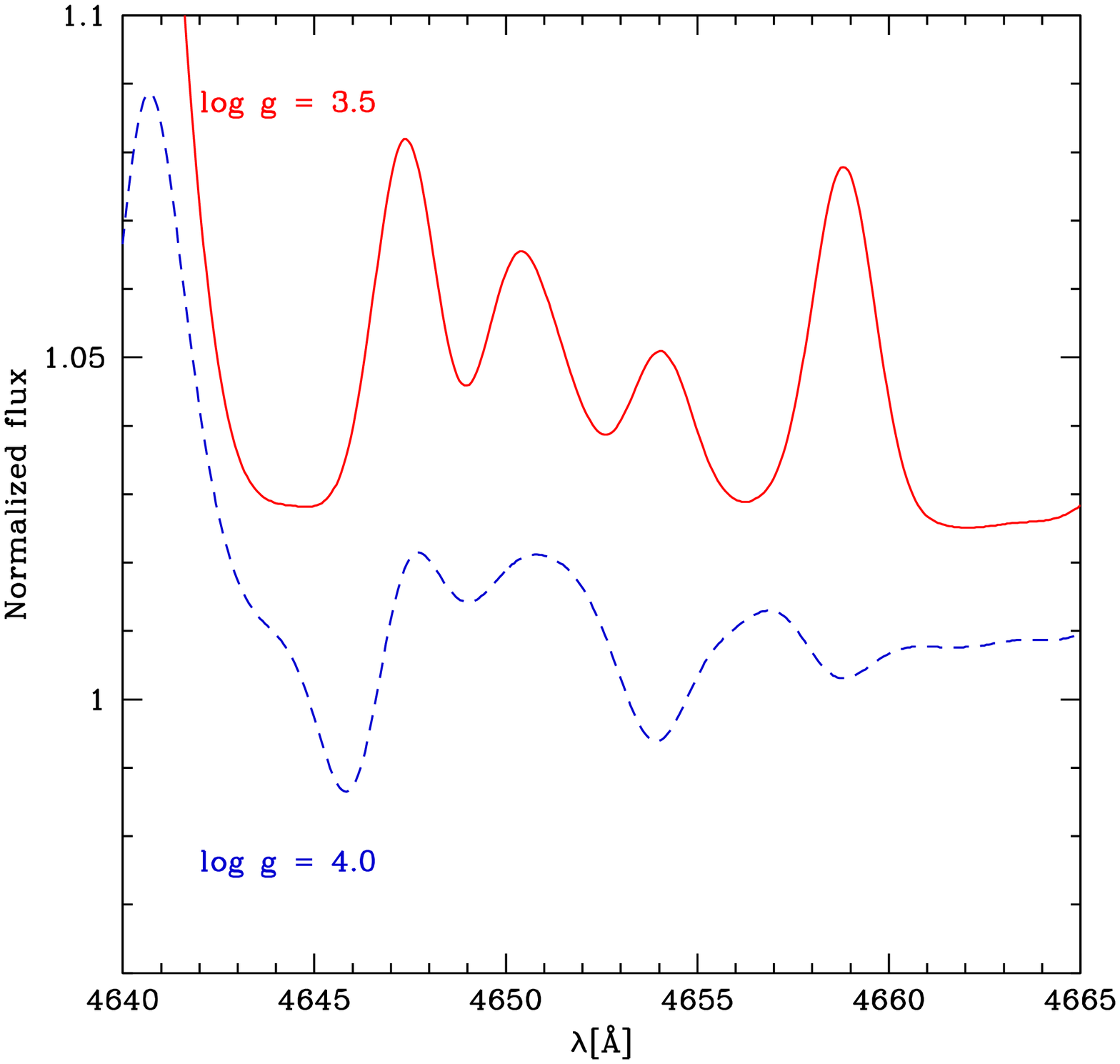}}
     \hspace{0.1cm}
     \subfigure[]{
%          \label{fig2}
          \includegraphics[width=.45\textwidth]{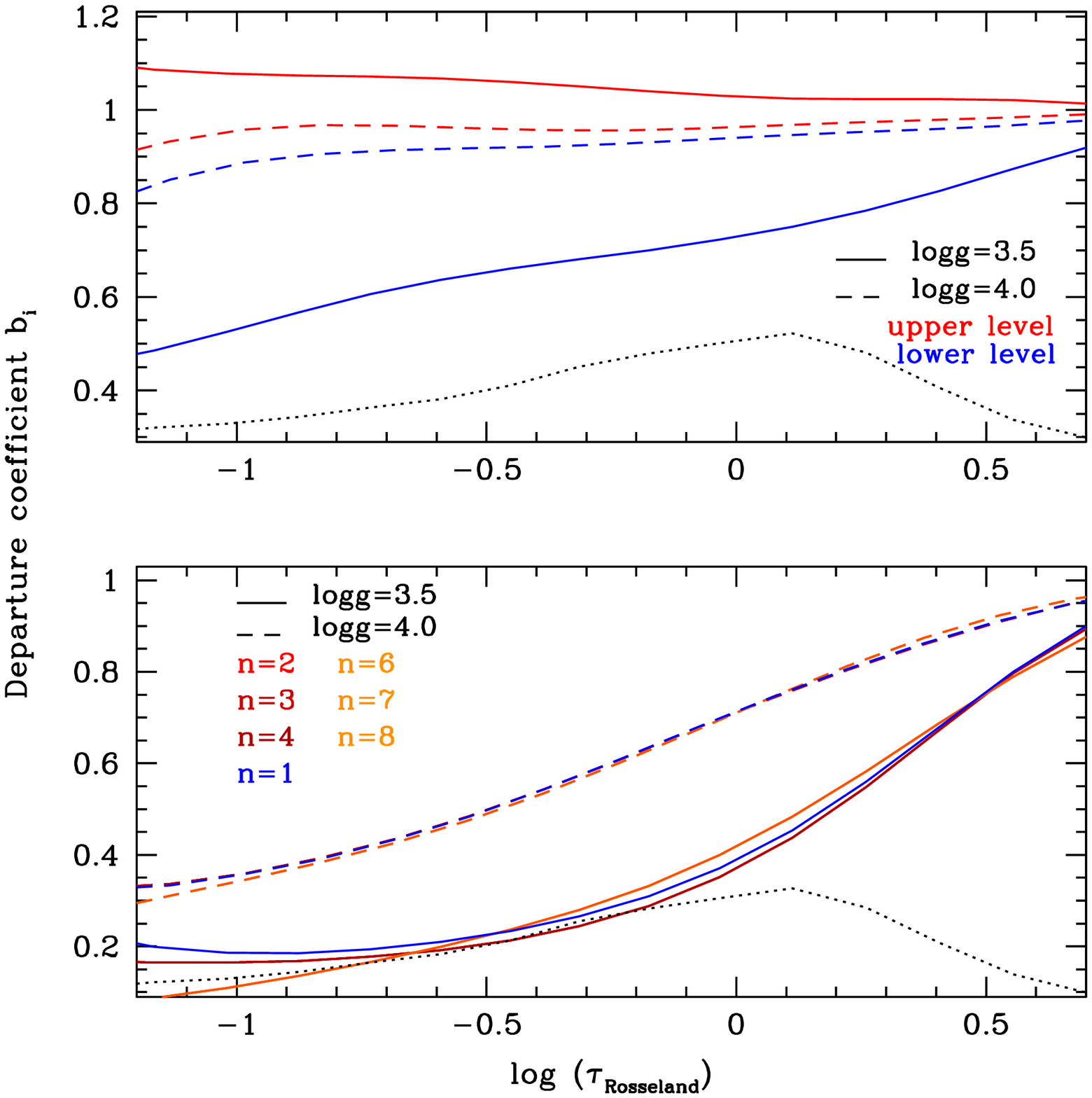}}
     \caption{Effect of \logg\ on \ion{C}{iii} 4647--50--51. \textit{Left panel}: the red line is a model with \logg\ = 3.5; the blue line is the same model with \logg\ = 4.0. The effective temperature is 42000 K. \textit{Right panel}: departure coefficient of the upper and lower levels of \ion{C}{iii} 4647--50--51 (upper panel) and of levels 2s$^2$~$^1$S (n=1), 2s2p~$^3$P$^{\scriptsize \rm o}$ (n=2,3,4), 2p2p~$^3$D (n=6,7,8) (lower panel) for \logg\ = 3.5 (solid line) and \logg\ = 4.0 (dashed line), plotted as a function of Rosseland opacity. The dotted line is the \ion{C}{iii} 4650 formation region.}
     \label{eff_logg_4650_t42}
\end{figure*}

%%%%%%%%%%%%%%%%%%%%%%%%%%%%%%%%%%%%%
\subsubsection{Effect of \logg}

This mechanism (depopulation of \ltl\ by \ion{C}{iii} 538) explains the behavior seen in Fig.\ \ref{eff_logg_4650_t42}. The effect of \logg\ on \ion{C}{iii} 4647-50-51 in a model at 42000 K is shown. The reduction of the gravity turns the weak emission in the \logg\ = 4.0 model into a strong emission in the \logg\ = 3.5 model. The departure coefficients are shown in the right panel of Fig.\ \ref{eff_logg_4650_t42}. While the upper level has $b_{i}$ close to 1 in both models, the lower level has a departure coefficient much lower than 1 in the low gravity model. This increases the line source function, leading to strong emission in the low gravity model.

Given the Grotrian diagram seen in Fig.\ \ref{gro_triplet} one can conclude that the mechanism is similar to that of \ion{C}{iii} 884 for \ciii. The higher ionization of the low \logg\ model leads to a depopulation of the ground level (2s$^2$ $^1$S) and the 2s2p $^3$P$^{\scriptsize \rm o}$ to which it is collisional coupled. The lower population of 2s2p $^3$P$^{\scriptsize \rm o}$ enhances the \ion{C}{iii} 538  drain (the total radiative rate is positive) depopulating 2s3s $^3$S and driving \ion{C}{iii} 4647-50-51 into emission. Hence, \ion{C}{iii} 538 controls the population of the lower level of \ion{C}{iii} 4650 and its morphology (absorption or emission)  at both high and low effective temperature. The situation for \ion{C}{iii} 4650 is clearer than for \ciii, since the only line affecting significantly its morphology is \ion{C}{iii} 538. In particular, \ion{C}{iii} 818, connected to \ltu\ (upper level of \ion{C}{iii} 4650), has no impact on the appearance of \ion{C}{iii} 4650 (Fig.\ \ref{eff_ciiilines_4650}). The reason is related to a quirk of the atomic physics: the Einstein coefficient $A$ for the 818 transition is only $2.3 \times 10^6$ compared with $7.2\times 10^7$ for \ion{C}{iii} 4650 and $4.1 \times 10^8$ for \ion{C}{iii} 538.

As for \ciii, we also tested the effect of low temperature dielectronic recombination 
and found out that its role was marginal. The line profiles show slightly more emission (or less absorption) when LTDR is included, but the changes are minor compared to the effects of UV lines highlighted above. In O stars, LTDR is not responsible for the strength of \ion{C}{iii} 4650.

%%%%%%%%%%%%%%%%%%%%%%%%%%%%%%%%%%%%%%%%%%%%%%%%%%%%%%%%%%%%%%%%%%%%%%%%%%%%%%%%%%%%%%%%%%%%%%%%%%%%%%
\subsection{Line blanketing effects}

%%%%%%%%%%%%%%%%%%%%%%%%
\subsubsection{Microturbulence and metallic lines}

As for \ciii\ we find that  microturbulence  and line blanketing effects influence the morphology of \ion{C}{iii} 4650.
In particular, an increase of \vturb\ lead to a significant strengthening of the absorption. Inspection of our line list also indicates that several metallic lines are located near the \ion{C}{iii} 538 transition -- the only UV line that significantly influences the strength of \ion{C}{iii} 4650.

\begin{figure}[]
\centering
\includegraphics[width=9cm]{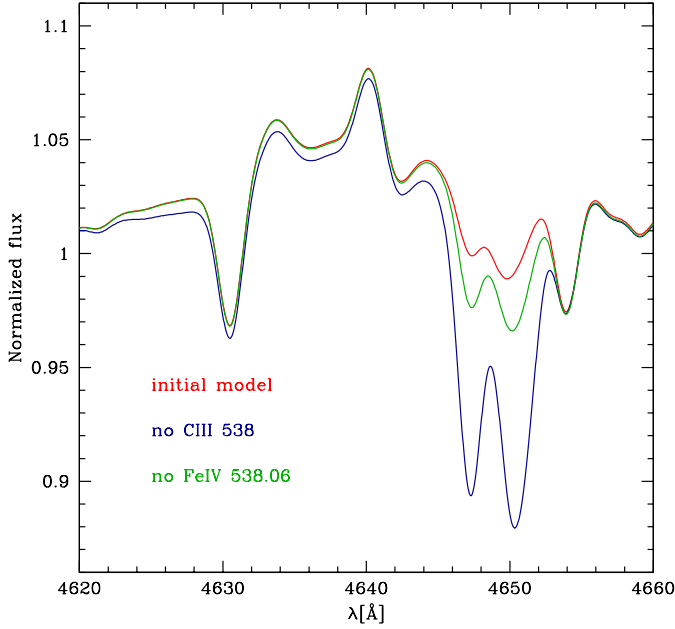}
\caption{Synthetic spectra around the \ion{C}{iii} 4650 triplet in models with \teff\ = 30500K. The red line is the initial model. The blue line is the same model in which we have removed the \ion{C}{iii} 538 \AA\ line. The green model is the same as the initial model in which the \ion{Fe}{iv} 538.057 \AA\ line has been removed.}
\label{eff_538}
\end{figure}

We tested the influence of the following lines on the strength of \ion{C}{iii} 4650: \ion{Fe}{iv} 538.044, \ion{Fe}{iv} 538.057, \ion{Fe}{vi} 538.097, \ion{Fe}{vi} 538.112, \ion{Fe}{iv} 538.176, \ion{Fe}{iv} 538.265, \ion{Fe}{vi} 538.285, \ion{O}{iii} 538.320, \ion{Ar}{iii} 538.348. Only \ion{Fe}{iv} 538.057 leads to a change of the \ion{C}{iii} 4650 profile -- its absence translates into a ``small'' increase in the absorption (Fig.~\ref{eff_538}). While the effect of \ion{Fe}{iv} 538.057 is important \ion{C}{iii} 538 has a much larger impact. The \ion{Fe}{iv} 538.057 has a negative total radiative rate and consequently assists in the depopulation of the \ltl\ state, like \ion{Fe}{iv} 574.232 for the upper level of \ciii.

%%%%%%%%%%%%%%%%%%%%%%%%
\subsubsection{Metallicity effects}

\begin{figure}[]
\centering
\includegraphics[width=9cm]{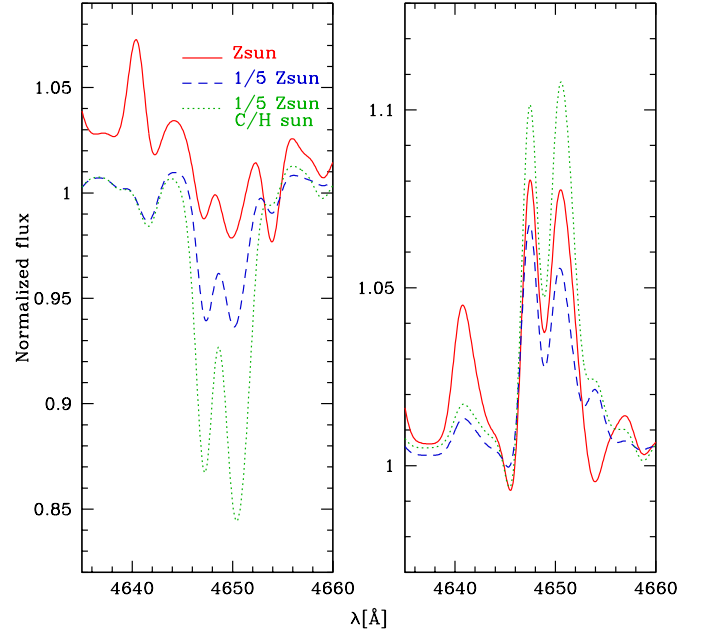}
\caption{Effect of metallicity on the \ion{C}{iii} 4650 triplet at 30500 K (left) and 42000 K (right). The solid line is a solar metallicity model. The dashed line is a model with $Z = 1/5^{th} Z_{\odot}$. The dotted line is the same model as the dashed line model but with a solar C/H.}
\label{eff_Z_4650}
\end{figure}

Fig.\ \ref{eff_Z_4650} shows the effect of metallicity on \ion{C}{iii} 4650 at \teff\ = 30500 K (left) and 42000 K (right). 
At low effective temperature, a global reduction of the metallicity translates into a stronger absorption. If the carbon content is kept at the solar value, the absorption is further strengthened, simply because there is more carbon and thus more \ion{C}{iii} in the atmosphere. The increase of the absorption due the reduction of Z can be understood as follows. We have seen in Fig.\ \ref{eff_538} that removing  \ion{C}{iii} 538 and \ion{Fe}{iv} 538.057 lead to an increase of the \ion{C}{iii} 4650 absorption. A reduction of Z corresponds to a reduction of the strength of \ion{C}{iii} 538 and \ion{Fe}{iv} 538.057. Consequently, a lower Z implies a stronger \ion{C}{iii} 4650 absorption. 

For \teff\ = 42000 K, the behavior of \ion{C}{iii} 4650 is the same as that of \ciii\ (see Appendix \ref{ap_Z_t42}). Reducing Z produces a weaker emission. But keeping C/H at the solar value and reducing the abundance of all other elements lead to a \ion{C}{iii} 4650 emission stronger than at solar metallicity. The reason is that the condition for emission are more favorable at low Z. In particular, the ionization is higher, draining the population of \ltl\ significantly even if \ion{C}{iii} 538 is weaker at low metallicity.

\begin{figure}[]
\centering
\includegraphics[width=9cm]{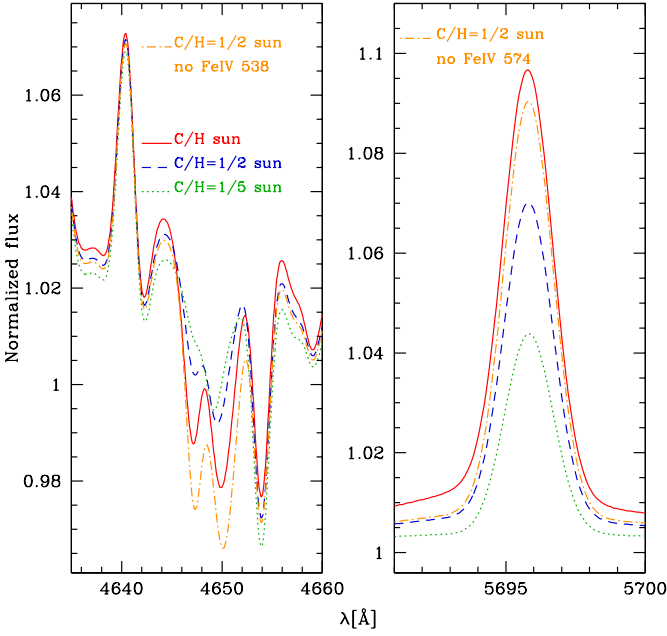}
\caption{Effect of C/H on the intensity of \ion{C}{iii} 4650 (left) and \ciii\ (right) at \teff\ = 30500 K. In both panels, the solid line is a solar metallicity model. The dashed (dotted) line is a model with C/H = 1/2 (1/5) C/H$_{\odot}$, all other elements having the solar abundances. In the left (right) panel, the dot--dashed line is the C/H = 1/2 C/H$_{\odot}$ model in which the \ion{Fe}{iv} 538.057 (574.232) line has been removed. }
\label{eff_ConH}
\end{figure}

As highlighted in the introduction, \ion{C}{iii} 4650 and \ciii\ are usually the strongest \ion{C}{iii} lines in O stars optical spectra. It is tempting to use them to perform carbon abundance determinations. However, we have seen that both lines critically depends on metallic lines the characteristics of which are not accurately known. We illustrate in Fig.\ \ref{eff_ConH} how such uncertainties can affect carbon abundance determinations. Removing \ion{Fe}{iv} 574.232 in a model with C/H = 1/2 C/H$_{\odot}$ leads to a \ciii\ profile almost as strong as in the solar C/H model (left panel). For \ion{C}{iii} 4650, the effect of removing \ion{Fe}{iv} 538.057 is even larger since the absorption in the C/H = 1/2 C/H$_{\odot}$ without this iron line is stronger than the absorption in the solar C/H model. In conclusion, using \ion{C}{iii} 4650 and \ciii\ to determine the carbon content of O stars is very risky, and quoting an uncertainty of less than a factor of two in the derived C/H using these lines is unrealistic.

%%%%%%%%%%%%%%%%%%%%%%%%%%%%%%%%%%%%%%%%%%%%%%%%%%%%%%%%%%%%%%%%%%%%%%%%%%%%%%%%%%%%%%%%%%%%%%%%%%%%%%%%%%%%%%%%%%%%%%%%%%%%%%%
%%%%%%%%%%%%%%%%%%%%%%%%%%%%%%%%%%%%%%%%%%%%%%%%%%%%%%%%%%%%%%%%%%%%%%%%%%%%%%%%%%%%%%%%%%%%%%%%%%%%%%%%%%%%%%%%%%%%%%%%%%%%%%%
%%
\section{Wind effects}
\label{s_wind}

We devote a special section to the effects of winds on the morphology of the \ion{C}{iii} optical lines. Details are also given in Appendix \ref{ap_wind}. In Fig.\ \ref{spec_wind}, we show the effects of changes in the mass loss-rate at 30500\,K and 42000\,K. The effects are complex, sometimes counter--intuitive, and do not change monotonically with mass-loss rate. The changes show that the optical lines cannot be modeled without modeling the wind, and that an accurate estimate of the mass-loss is required. Because of clumping, obtaining an accurate estimate of the mass-loss rate is not easy. Because mass-loss rates strongly correlate with luminosity, we expect the \ion{C}{iii} optical lines (at fixed \teff\ and \logg) to correlate with luminosity.

\begin{figure*}[t]
     \centering
     \subfigure[]{
%          \label{fig1}
          \includegraphics[width=.45\textwidth]{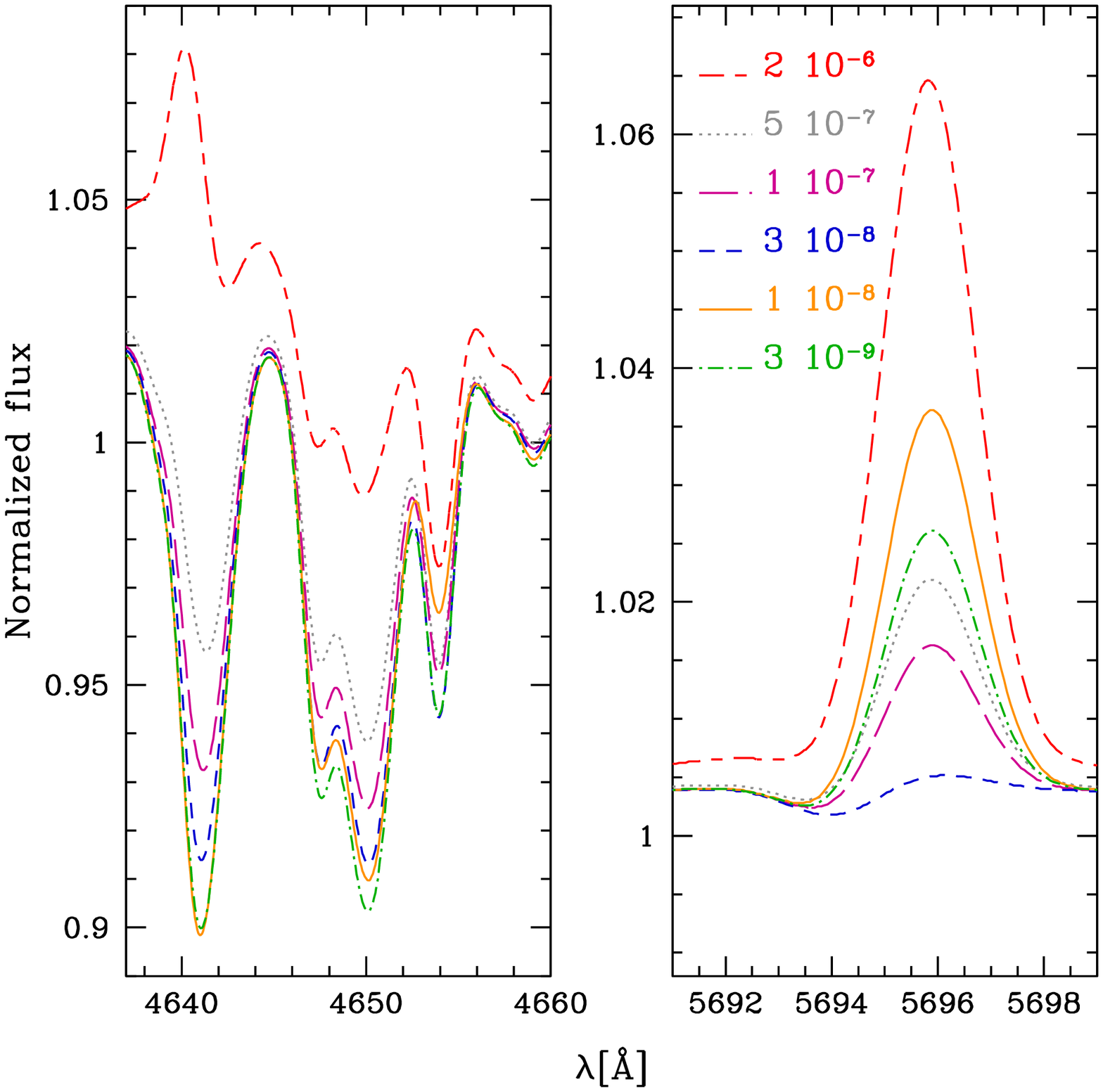}}
     \hspace{0.1cm}
     \subfigure[]{
%          \label{fig2}
          \includegraphics[width=.45\textwidth]{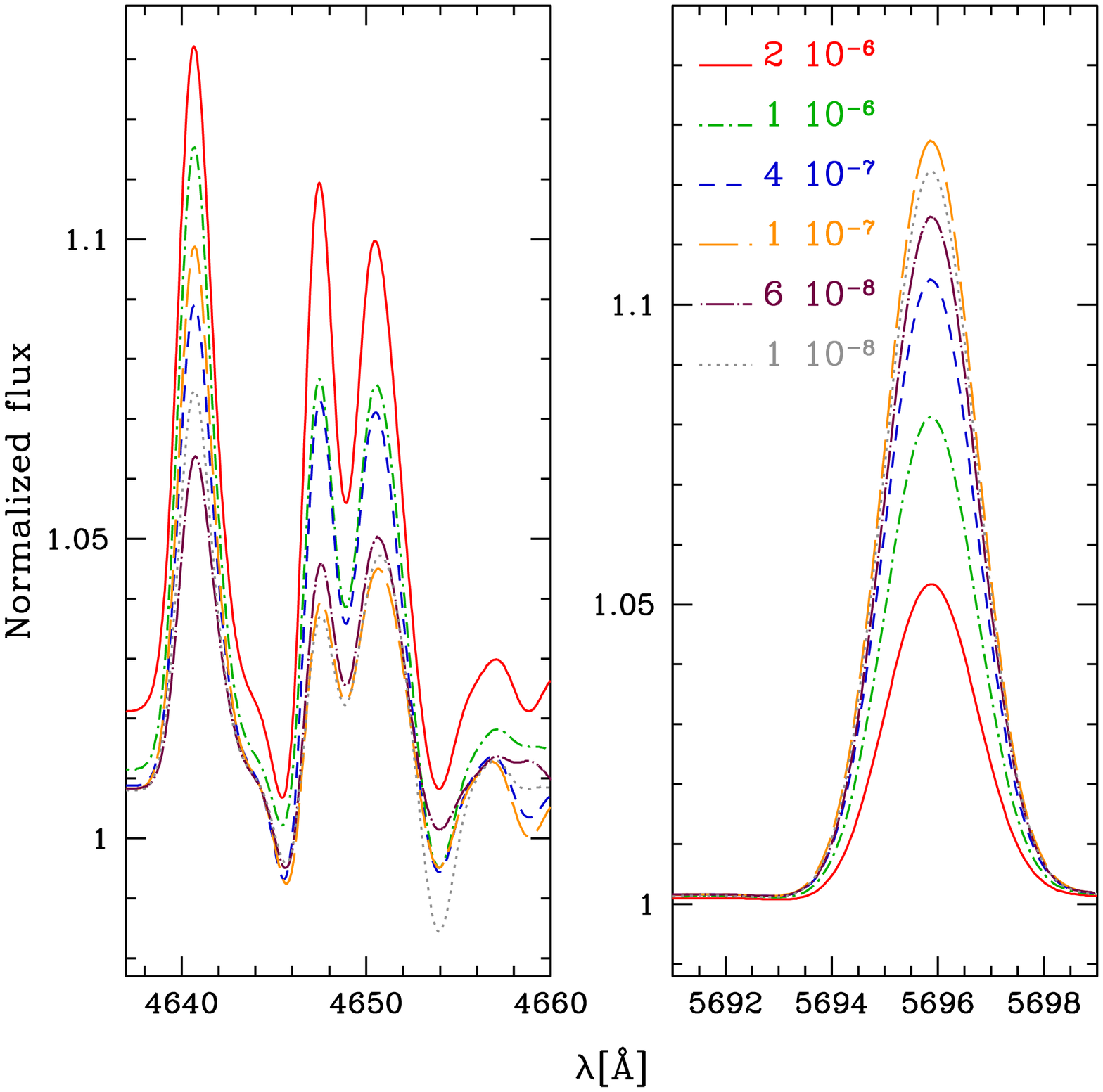}}
     \caption{Effects of a change in the mass loss rate on \ion{C}{iii} 4650 and \ciii\ in models with \teff\ = 30500 K (left) and \teff\ =42000 K (right). The mass loss rates (in \myr) are indicated in the \ciii\ panels.}
     \label{spec_wind}
\end{figure*}

Let us first examine \ciii. At \teff\ = 30500 K, starting from the low mass loss rate model ($3 \times\ 10^{-9}$ \myr), we see that the emission line profile first strengthens when \mdot\ increases. A further increase of \mdot\ from $10^{-8}$ to $3 \times\ 10^{-8}$ \myr\ translates into a change from emission to weak absorption. Above $3 \times\ 10^{-8}$ \myr, the increase of \mdot\ resumes to produce stronger emission. Thus, the behavior of the \ciii\ line strength/morphology with increasing mass loss rate is not monotonic. At higher effective temperature, there is  a clear anti-correlation of strength of \ciii\ emission with mass loss rate above $10^{-7}$ \myr. Below that value, the emission stregnth is almost at the level of the  $10^{-7}$ \myr\ model. 

For \ion{C}{iii} 4650 at low \teff, an increase of the mass loss rate is associated with a decrease of the absorption and/or an increase of the emission is seen. This is what is commonly observed in wind sensitive lines such as H$\alpha$. In the 42000\,K model, below $10^{-7}$ \myr, \ion{C}{iii} 4650 has a rather constant emission level, with slight fluctuations. Above $10^{-7}$ \myr, the emission increases with mass loss rate.

A detailed explanation of the observed behavior is provided in Appendix \ref{ap_wind}. Basically, it is related to the influence of UV lines on the populations of the upper levels. As we increase the mass-loss rate we increase the velocity at a fixed Rosseland optical depth, and this alters the radiation transfer in the UV lines. For example, at higher velocities the lines desaturate earlier which can enhance both photon escape and photon pumping. Which of the two processes dominates depends on both the radiation field in the lines neighborhood and the line under consideration. These two processes have opposite effects on the strength of an optical line -- photon escape in the UV line enhances optical emission (weakens absorption) while photon pumping enhances absorption (weakens emission) when the UV line's upper state is the lower level of the optical transition. The converse occurs when the UV lines is connected to the upper state of the optical transition. The behavior of \ciii\ at low mass loss rates is complex and not straightforward.

%%%%%%%%%%%%%%%%%%%%%%%%%%%%%%%%%%%%%%%%%%%%%%%%%%%%%%%%%%%%%%%%%%%%%%%%%%%%%%%%%%%%%%%%%%%%%%%%%%%%%%%%%%%%%%%%%%%%%%%%%%%%
%%%%%%%%%%%%%%%%%%%%%%%%%%%%%%%%%%%%%%%%%%%%%%%%%%%%%%%%%%%%%%%%%%%%%%%%%%%%%%%%%%%%%%%%%%%%%%%%%%%%%%%%%%%%%%%%%%%%%%%%%%%%
\section{Concluding remarks}
\label{s_conc}

We have investigated the formation processes of \ciii\ and \ion{C}{iii} 4650 in the atmosphere of O stars. We have used the non--LTE atmosphere code CMFGEN to perform simulations at two effective temperatures (30500 K and 42000 K) corresponding to late and early spectral types. Our results can be summarized as follows:

\begin{itemize}

\item[$\bullet$] For wind densities typical of O stars, the formation of \ciii\ is controlled by several UV transitions: \ion{C}{iii} 386, \ion{C}{iii} 574 and \ion{C}{iii} 884. The latter is crucial at low \teff: when it is removed from models, \ciii\ switches from emission to absorption. \ion{C}{iii} 574 and \ion{C}{iii} 386 play the same role at high \teff.

\item[$\bullet$] \ion{C}{iii} 884 acts as a drain on the \lsl\ population (lower level of \ciii). When it is removed, \lsl\ is more populated and emission is reduced/suppressed. \ion{C}{iii} 574 drains the \lsu\ population and \ion{C}{iii} 386 populates \lsl. When they are removed, \lsu\ is more populated and \lsl\ is depopulated, favoring emission.

\item[$\bullet$] Lower gravity models have larger ionizing flux. As a consequence, the lower level of \ion{C}{iii} 884 (2p2p~$^1$F$^{\scriptsize \rm o}$), which is collisionally coupled to the ground level, is depopulated at low \logg. The drain of \lsl\ by \ion{C}{iii} 884 is thus increased. This leads to an increase of \ciii\ emission when \logg\ decreases at 30500 K. At \teff\ = 42000 K, the effect of \logg\ is opposite and is due to the role of \ion{C}{iii} 386 in populating the lower level of \ciii.

\item[$\bullet$] The strength of \ciii\ is very sensitive to the action of a few metallic lines on the radiative transfer near 386, 574 and 884 \AA. In particular, \ion{Fe}{iv} 386.262 and \ion{Fe}{iv} 574.232, when removed from models, lead to a stronger \ciii\ emission. If \ion{S}{v} 884.531 is removed, the \ciii\ emission is reduced. The reason is the overlap of these features with the key \ion{C}{iii} lines controlling the shape of \ciii. As an example, \ion{Fe}{iv} 574.232 extracts photons from \ion{C}{iii} 574 leading to a lower population of \lsu. Consequently, \ion{Fe}{iv} 574.232 limits the \ciii\ emission. When it is removed, \lsu\ is more populated and the \ciii\ emission is stronger.

\item[$\bullet$] The formation of \ion{C}{iii} 4650 is controlled by radiative transfer at 538 \AA. The \ion{C}{iii} 538 line drains the population of \ltl, the lower level of \ion{C}{iii} 4650. When \ion{C}{iii} 538 is removed, \ltl\ is more populated and \ion{C}{iii} 4650 switches from emission to absorption, or shows a stronger absorption.

\item[$\bullet$] \ion{Fe}{iv} 538.057 enhances photon escape in the \ion{C}{iii} 538 line which helps to reduce the population of \ltl, and thus weakens the \ion{C}{iii} 4650 absorption (or enhances the emission). 

\item[$\bullet$] Low temperature dielectronic recombinations have a limited effect on the appearance of \ciii\ and \ion{C}{iii} 4650 in O stars.

\item[$\bullet$] When the mass loss rate increases, a growing fraction of the \ion{C}{iii} 4650 and \ciii\ lines is formed in the wind, leading to a larger emission. In addition, as \mdot\ gets higher, the formation region of the \ion{C}{iii} lines (UV and optical) shifts to higher velocities. The associated Doppler shifts lead to desaturation of the key \ion{C}{iii} UV lines. This affects the level populations and consequently the intrinsic \ion{C}{iii} 4650 and \ciii\ line profiles. At low mass-loss rates, the change in line profile with a change in the mass-loss rate is not monotonic.

\item[$\bullet$] Metallicity effects are a complex interplay between \ion{C}{iii} UV lines, metallic \ion{Fe}{iv} and \ion{S}{v} lines, ionization, winds and abundance effects. There is no simple way of predicting how \ion{C}{iii} 4650 and \ciii\ will react to a change of metallicity or abundance of specific elements. 

\end{itemize} 

In view of our results, it seems highly risky to derive accurate carbon abundances from \ion{C}{iii} 4650 and \ciii\ in O star atmospheres. Uncertainties of at least a factor of two on C/H are expected just because of the uncertainty on atomic data and wavelengths of metallic lines. Priority should be given to other carbon lines, such as the \ion{C}{iii} 2s4f\,$^3$F$^{\scriptsize \rm o}$-2s5g\,$^3$G multiplet at 4070\,\AA, or even UV lines (\ion{C}{iv} 1164 and \ion{C}{iii} 1176) provided the wind is not too strong and they are mainly photospheric. 

The current results can be used to try to understand the behavior of \ion{C}{iii} 4650 in Of?p stars, many (all?) of which are magnetic  \citep{donati06,hubrig08,martins10,wade11,wade191612}. Of?p stars show periodic variability, especially in \ion{C}{iii} 4650 emission which can be as strong as in the neighbouring \ion{N}{iii} lines before almost vanishing. H$_{\alpha}$ shows the same variability and is strongest when \ion{C}{iii} is at its maximum emission strength. \citet{sund12} demonstrated that the changes in H$_{\alpha}$ emission of HD~191612 could be understood as a result of viewing the dynamic magnetosphere from different angles.

Due to wind confinement by the magnetic field, an overdense region is created in the magnetic equator. If the star is seen from the magnetic pole, a large surface of the overdense region is observed and the H$_{\alpha}$ emission is stronger than if the star is seen from the magnetic equator. In the present study, we have seen that \ion{C}{iii} 4650 can show emission if the wind contribution becomes dominant, and/or if for some reason the drain effect of \ion{C}{iii} 538 is increased so that \ltl\ is depopulated. If we observe HD~191612 from the magnetic pole, the wind velocity at a given radius is larger than in the magnetic equator, due to confinement (see Fig.\ 2 of Sundqvist et al.\ 2012). The role of the wind is thus stronger, favoring \ion{C}{iii} 4650 emission. In addition, since the atmosphere density is lower in the magnetic pole direction, we can expect a larger ionizing flux due to reduced opacity. As seen in our study, this favors the drain of \ltl\ through \ion{C}{iii} 538, and thus \ion{C}{iii} 4650 emission. Hence, the existence of an equatorial overdensity could lead to the observed \ion{C}{iii} 4650: more emission when the star is seen pole--on, less emission when it is seen equator--on, just as H$_{\alpha}$. Of course, this needs to be tested by detailed 2D radiative transfer calculation, but the present results provide a good working frame.

%%%%%%%%%%%%%%%%%%%%%%%%%%%%%%%%%%%%%%%%%%%%%%%%%%%%%%%%%%%%%%%%%%%%%%%%%%%%%%%%%%%%%%%%%%%%%%%%%%%%%%%%%%%%%%%%%%%%%%%%%%%%
%%%%%%%%%%%%%%%%%%%%%%%%%%%%%%%%%%%%%%%%%%%%%%%%%%%%%%%%%%%%%%%%%%%%%%%%%%%%%%%%%%%%%%%%%%%%%%%%%%%%%%%%%%%%%%%%%%%%%%%%%%%%
\begin{acknowledgements}
FM acknowledges support from the Agence Nationale de la Recherche. DJH acknowledges support from STScI theory grants HST-AR-11756.01.A and HST-AR-12640.01. STScI is operated by the Association of Universities for Research in Astronomy, Inc., under NASA contract NAS5-26555.
\end{acknowledgements}

%%%%%%%%%%%%%%%%%%%%%%%%%%%%%%%%%%%%%%%%%%%%%%%%%%%%%%%%%%%%%%%%%%%%%%%%%%%%%%%%%%%%%%%%%%%%%%%%%%%%%%%%%%%%%%%%%%%%%%%%%%%%
%%%%%%%%%%%%%%%%%%%%%%%%%%%%%%%%%%%%%%%%%%%%%%%%%%%%%%%%%%%%%%%%%%%%%%%%%%%%%%%%%%%%%%%%%%%%%%%%%%%%%%%%%%%%%%%%%%%%%%%%%%%%
\bibliographystyle{aa}
\bibliography{biblio.bib}

%%%%%%%%%%%%%%%%%%%%%%%%%%%%%%%%%%%%%%%%%%%%%%%%%%%%%%%%%%%%%%%%%%%%%%%%%%%%%%%%%%%%%%%%%%%%%%%%%%%%%%%%%%%%%%%%%%%%%%%%%%%%
%%%%%%%%%%%%%%%%%%%%%%%%%%%%%%%%%%%%%%%%%%%%%%%%%%%%%%%%%%%%%%%%%%%%%%%%%%%%%%%%%%%%%%%%%%%%%%%%%%%%%%%%%%%%%%%%%%%%%%%%%%%%
%\Online
\begin{appendix}

\section{Metal line atomic data}
\label{ap_met_line}

\begin{table}[h]
\begin{center}
\caption{Elements and number of levels (super--levels) included in the CMFGEN models.} \label{tab_elmts}
\begin{tabular}{lcc}
\hline
              & \teff\ = 30500 K & \teff\ = 42000 K \\
\hline   
\ion{H}{i}    &  30 (30)          &  30 (30) \\
\ion{He}{i}   &  69 (69)          &  69 (69) \\
\ion{H}{i}    &  30 (30)          &  30 (30) \\
\ion{C}{ii}   &  39 (21)          &  -- \\
\ion{C}{iii}  &  243 (99)         &  243 (99) \\
\ion{C}{iv}   &  64 (64)          &  64 (64) \\
\ion{N}{ii}   &  105 (59)         &  -- \\
\ion{N}{iii}  &  287 (57)         &  287 (57) \\
\ion{N}{iv}   &  70 (44)          &  70 (44) \\
\ion{N}{v}    &  49 (41)          &  49 (41) \\
\ion{O}{ii}   &  274 (155)        &  -- \\
\ion{O}{iii}  &  104 (36)         & 104 (36) \\
\ion{O}{iv}   &  64 (30)          & 64 (30) \\
\ion{O}{v}    &  56 (32)          & 56 (32) \\
\ion{Ne}{ii}  &  48 (14)          & 48 (14) \\
\ion{Ne}{iii} &  71 (23)          & 71 (23) \\
\ion{Ne}{iv}  &  52 (17)          & 52 (17) \\
\ion{Si}{iii} &  50 (50)          & 50 (50) \\ 
\ion{Si}{iv}  &  66 (66)          & 66 (66) \\ 
\ion{S}{iii}  &  78 (39)          & 78 (39) \\
\ion{S}{iv}   &  108 (40)         & 108 (40) \\
\ion{S}{v}    &  144 (37)         & 144 (37) \\
\ion{Ar}{iii} &  138 (24)         & 346 (32) \\
\ion{Ar}{iv}  &  102 (30)         & 382 (50) \\
\ion{Ar}{v}   &  29 (14)          & 376 (64) \\
\ion{Ca}{iii} &  88 (28)          & 232 (44) \\
\ion{Ca}{iv}  &  72 (19)          & 378 (43) \\
\ion{Fe}{iii} &  607 (65)         & 607 (65) \\
\ion{Fe}{iv}  &  1000 (100)       & 1000 (100) \\
\ion{Fe}{v}   &  1000 (139)       & 1000 (139) \\
\ion{Fe}{vi}  &  1000 (59)        & 1000 (59) \\
\ion{Fe}{vii} & --                & 245 (40) \\
\ion{Ni}{iii} &  150 (24)         & 150 (24) \\
\ion{Ni}{iv}  &  200 (36)         & 200 (36) \\
\ion{Ni}{v}   &  183 (46)         & 183 (46) \\
\ion{Ni}{vi}  &  182 (40)         & 182 (40) \\
\ion{Ni}{vii} & --                & 67 (20) \\
\hline
\end{tabular}
\end{center}
\end{table}

Table \ref{tab_elmts} provides the number of levels and super--levels used in our computations. Basic atomic data for the important transitions that influence C\,{\sc iii} 5696 and 4650 are listed in Table \ref{Tab_atomic}. There is some uncertainty in the oscillator strengths, and this must be regarded as a source of uncertainty when modeling the C\,{\sc iii} lines. For a comparison we provide an alternative f values that is available in the literature for each transition. Fortunately, for the transitions which significantly influence C\,{\sc iii} 5696 and 4650, the wavelengths and/or energy levels for the relevant transition are known.  The wavelengths of most of the other transitions tested are (very) uncertain, but potentially could have influenced the modeling because of line overlap. A more worrisome concern is there may be some other important transition which overlaps with one of the key UV transitions, and which has  a wrong wavelength (because it has not been measured in the laboratory) or was omitted from the model atoms used in the model atmosphere. The work presented in the current paper, and that of \cite{NHP06_HeI}, highlights the need for accurate Fe wavelengths and oscillator strengths for modeling O star spectra.

\begin{table*}[]
\begin{center}
\caption{Basic line data for the important overlapping transitions}
\begin{tabular}{lllll}
\hline
 Species             & Transition      &         Wavelength(\AA)       &       f(Model)$^1$ & f(Alt)$^2$    \\ 
 \hline
Fe\,{\sc iv}    &  3d$^5$ $^4$F$_{5/2}$-3d$^4$($^3$D)4p $^4$D$^o_{5/2}$ & 574.232 & 0.00523 & 0.0131 \\
Fe\,{\sc v}     &  3d$^4$ $^5$D$_o$-3d$^3$($^4$F)4p $^5$D$^o_1$  &  386.262 & 0.0406 & 0.0341 \\
S\,{\sc v}       &  3s3d $^3$D$_3$ -3p3d $^3$D$^o_2$ &     884.531 & 0.0266 & 0.00745 \\
Fe\,{\sc vi}    &  3d$^5$ $^4$G$_{7/2}$- 3d$^4$($^3$H)4p $^4$G$^o_{7/2}$   & 538.057 &  0.121 & 0.0174 \\
\hline
\end{tabular}
\end{center}
\label{Tab_atomic}

\vskip 5pt
1. Model oscillator strengths are from R.L. Kurucz and B. Bell (1995 Atomic Line Data) Kurucz CD-ROM No. 23. Cambridge, Mass.: Smithsonian Astrophysical Observatory. Newer data are available at http://kurucz.harvard.edu/ \citep{Kur09_ATD} \\
2. Alternate f values for Fe are from \cite{BB95_FeIV, BB92_FeV,BB95_FeVI} while that for S\,{\sc v} is from
\cite{NIST}.
\end{table*}

%\end
%%%%%%%%%%%%%%%%%%%%%%%%%%%%%%%%%%%%%%%%%%%%%%%%%%%%%%%%
\section{Effect of metallic lines on \ion{C}{iii} 8500}
\label{s_8500}

The Grotrian diagram for the \ion{C}{iii} singlets (Fig.\ \ref{gro_singlet}) shows that the \ion{C}{iii} 8500 transition should also strongly depend on the efficiency of the \ion{C}{iii} 884 and \ion{C}{iii} 386 drains since \lsl\ is the upper level of all three transitions. Consequently, any modification of the \lsl\ population will change the \ion{C}{iii} 8500 morphology.

\begin{figure}
\centering
\includegraphics[width=9cm]{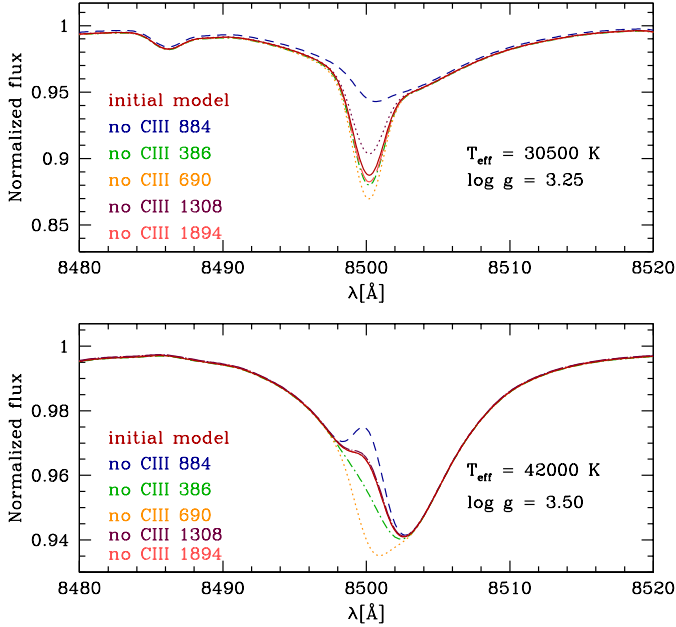}
\caption{Effect of \ion{C}{iii} UV lines on the appearance of \ion{C}{iii} 8500 for \teff\ = 30500 K (top panel) and 42000 K (lower panel). }
\label{ciii8500}
\end{figure}

In Fig.\ \ref{ciii8500} we show how this line reacts to the \ion{C}{iii} UV lines at \teff\ = 30500 K (upper panel) and \teff = 42000 K (lower panel). As expected, removing the drain from \ion{C}{iii} 884 translates into a weaker absorption at low \teff\ and to emission at high \teff. \ion{C}{iii} 386 populates \lsl\ at high \teff\ and thus removing it leads to absorption in \ion{C}{iii} 8500. \ion{C}{iii} 690 has also a strong drain effect on 2s3s $^1$S since removing it leads to a strong absorption. The \ion{C}{iii} 8500 transition is of special interest in the context of future observations of O stars by \textit{Gaia}, since it is basically the only non H/He line in the spectral window of the \textit{Radial Velocity Spectrometer} onboard the satellite. Understanding its properties is thus important to classify O stars observed by \textit{Gaia}.

%%%%%%%%%%%%%%%%%%%%%%%%%%%%%%%%%%%%%%%%%%%%%%%%%%%%%%%%
\section{Effect of metallicity on \ciii\ for \teff\ = 42000 K}
\label{ap_Z_t42}

\begin{figure}
\centering
\includegraphics[width=9cm]{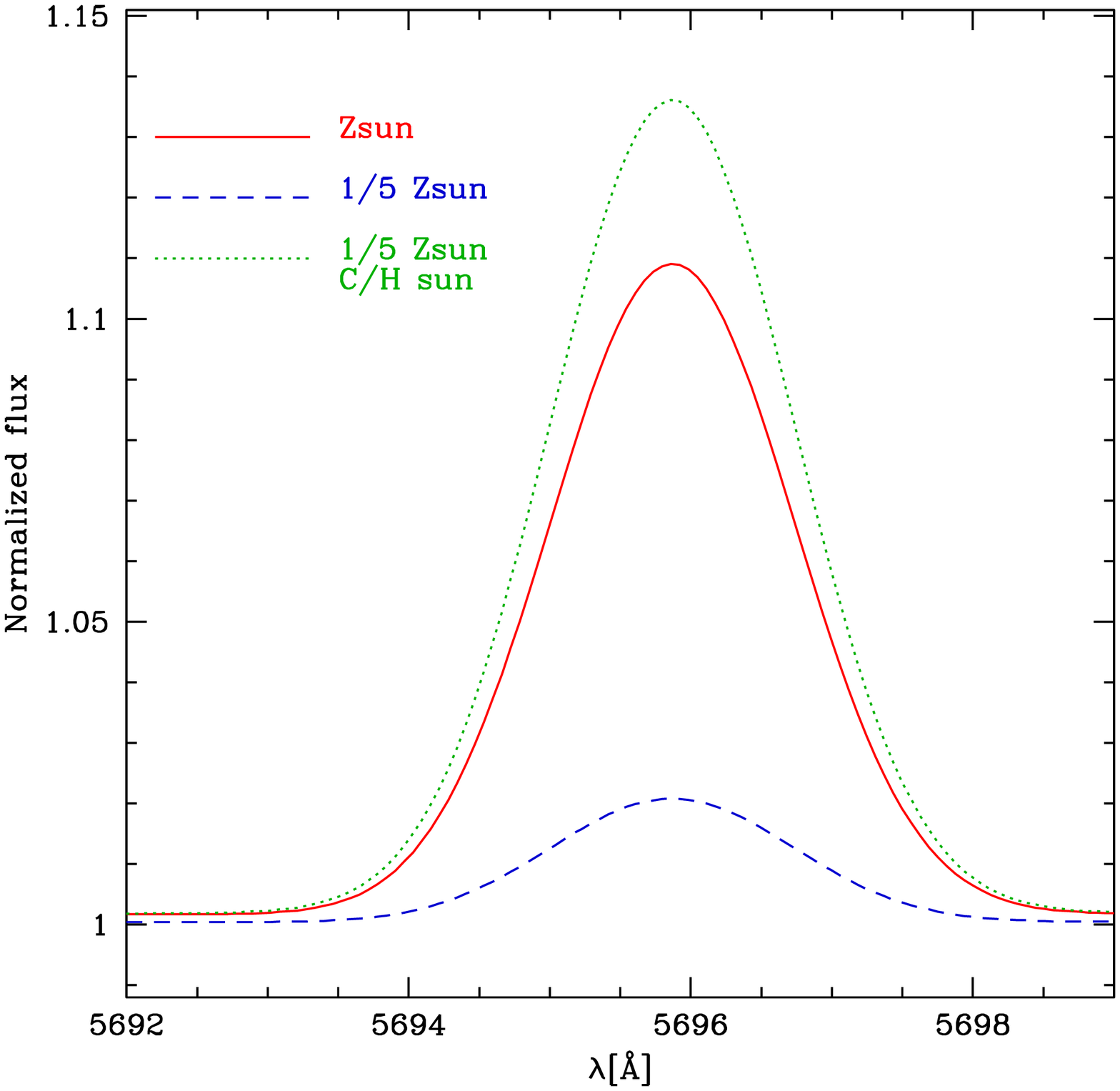}
\caption{Effect of metallicity on \ciii\ in the model with \teff\ = 42000 K. }
\label{spec_Z_t42}
\end{figure}

Fig.\ \ref{spec_Z_t42} presents the effect of a change of metallicity on \ciii\ in the model with \teff\ = 42000 K. We show a model with a global reduction of Z by a factor 5 (dashed line) and a model with the same reduction, but in which C/H is kept at the solar value (dotted line). The solid line is the solar metallicity model. A global reduction of Z (including a reduction of C/H) by a factor 5 leads to a significant reduction of the emission in \ciii. But a model with $Z =$ 1/5$^{th}$ Z$_{\odot}$ and a solar carbon content has a \ciii\ line slightly \textit{stronger} than the solar metallicity model. The higher ionizing flux at low Z implies a depopulation of \lsl\ and thus more emission. In addition, the reduced effect of \ion{Fe}{iv} 574.232, \ion{Fe}{v} 386.262 and \ion{S}{v} 884.531 tend to increase emission (see Fig.\ \ref{spec_felines}). All in all, the conditions are more favorable at low Z, explaining that for a similar C/H, the emission is stronger than at solar metallicity.

%%%%%%%%%%%%%%%%%%%%%%%%%%%%%%%%%%%%%%%%%%%%%%%%%%%%%%%%
\section{An explanation for the influence of the wind on optical \ciii\ lines}
\label{ap_wind}

In this section we provide an investigation of the mechanisms responsible for the trend observed in Fig.\ \ref{spec_wind} which shows the effect of mass loss on \ion{C}{iii} 4650 and \ciii. 

\begin{figure}[]
\centering
\includegraphics[width=9cm]{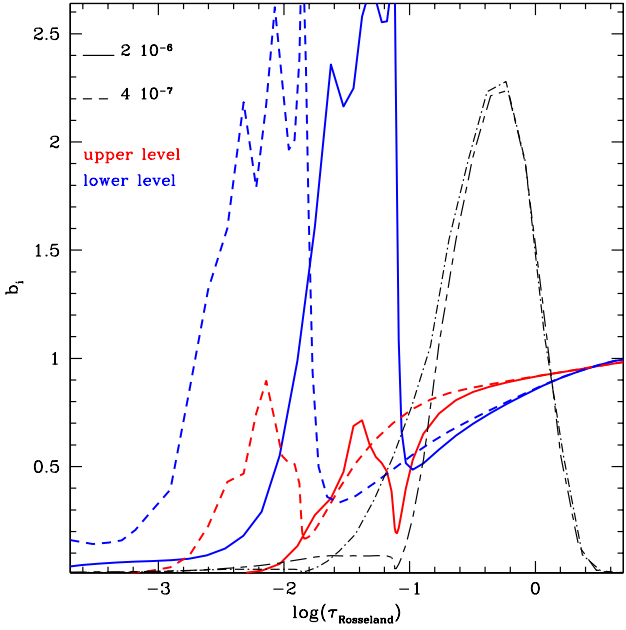}
\caption{Effects of mass loss rate on \lsl\ (blue) and \lsu\ (red) departure coefficients in a model with \teff\ = 42000 K. The dashed (solid) lines corresponds to a model with \mdot\ = $4 \times\ 10^{-7}$ ($2 \times\ 10^{-6}$) \myr. The dot--dashed (short dash -- long dashed) line is the line formation region in the low (high) \mdot\ model.}
\label{bi_wind_t42}
\end{figure}

We first focus on \ciii. Its behavior in the 42000 K model can be understood as follows. 
Fig.\ \ref{bi_wind_t42} shows the departure coefficients of the lower and upper levels of \ciii\ in a model with \mdot\ = $4 \times\ 10^{-7}$ \myr\ (dashed line) and $2 \times\ 10^{-6}$ \myr\ (solid line). In the low \mdot\ model, 88\%\ of the line is formed at $\log \tau_{\rm Rosseland} > -1.8$, corresponding to velocities smaller than 10 \kms. In that region, the lower level is always less populated than the upper level, a favorable condition for emission. In the high \mdot\ model, a significant fraction of \ciii\ (41\%) is formed at velocities higher than 10\kms ($\log \tau_{\rm Rosseland} < -1.2$). In the low velocity part of the formation region, the lower level is less populated than the upper level, but above 10 \kms, the lower level is much more populated. As a result, the global \ciii\ emission is weaker than in the low \mdot\ model. The behavior of \ciii\ with mass loss observed in the left panel of Fig.\ \ref{spec_wind} is thus a direct consequence of the strengthening of the wind.

\begin{figure*}[t]
     \centering
     \subfigure[]{
%          \label{fig1}
          \includegraphics[width=.45\textwidth]{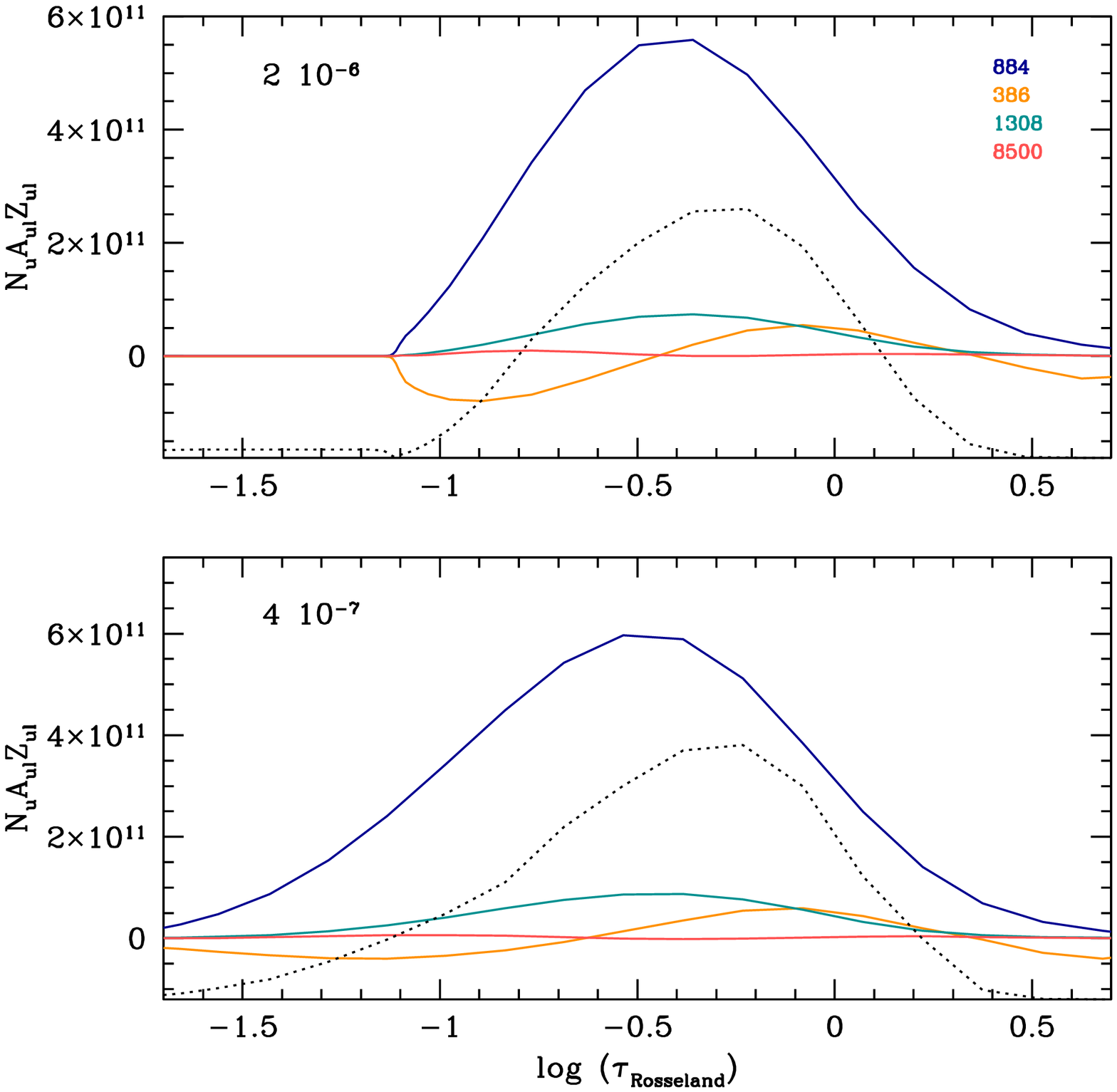}}
     \hspace{0.1cm}
     \subfigure[]{
%          \label{fig2}
          \includegraphics[width=.45\textwidth]{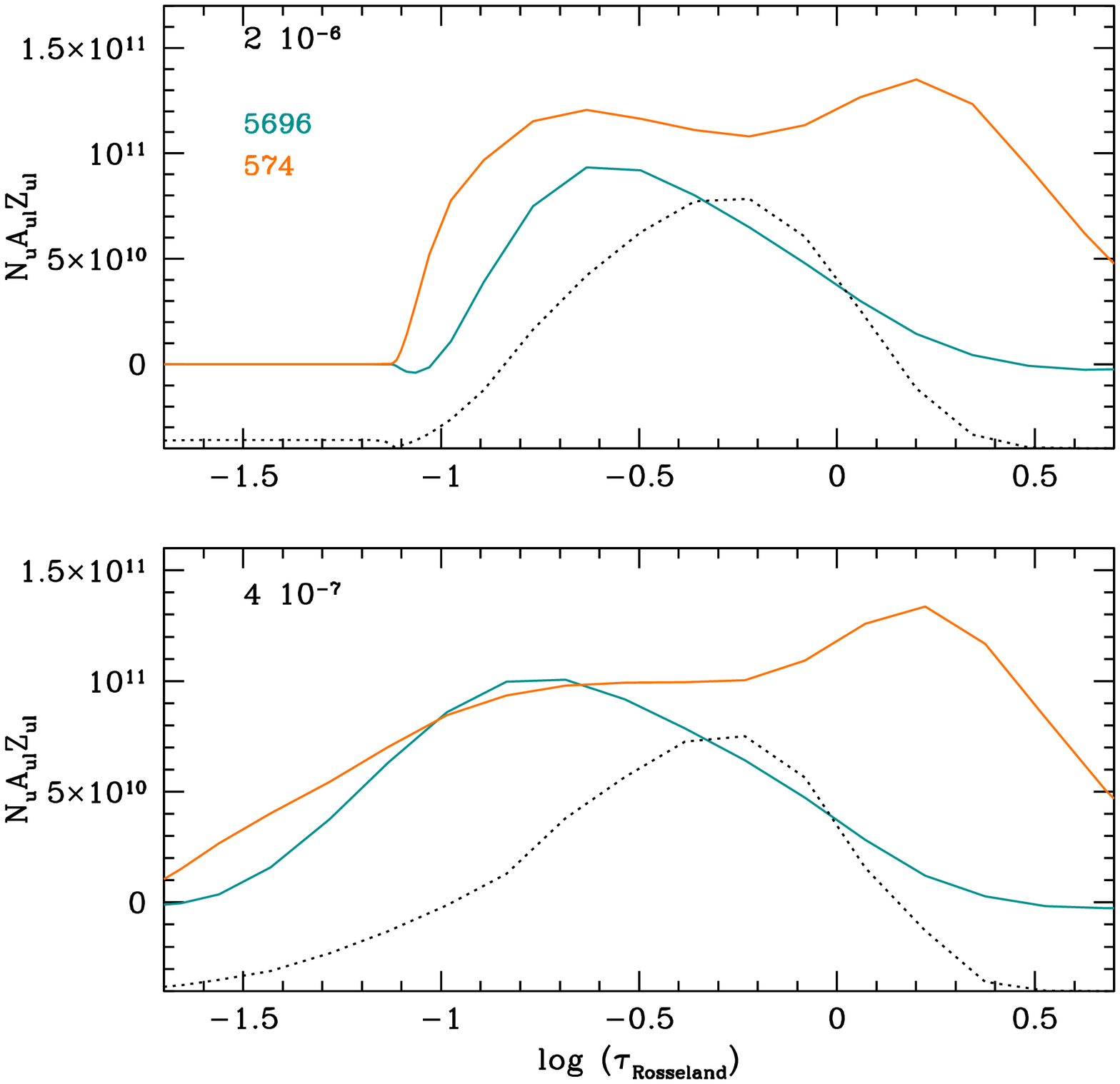}}
     \caption{Effect of mass loss rate on total radiative rates for lines from level \lsl\ (left) and level 2s3d1D (right). The models have \teff\ = 42000 K. The upper panels correspond to the model with \mdot\ = $2 \times\ 10^{-6}$ \myr, the lower panel to models with \mdot\ = $4 \times\ 10^{-7}$ \myr.}
     \label{rates_5696_wind_t42}
\end{figure*}

The behavior of \lsl\ and \lsu\ when the wind sets in is explained as follows. Fig.\ \ref{rates_5696_wind_t42} shows the total radiative rates in lines connecting the upper and lower levels of \ciii. The rates are shown for two mass loss rates discussed above ($4 \times\ 10^{-7}$ and $2 \times\ 10^{-6}$ \myr). When \mdot\ is increased, we see 1) that the rate in the \ion{C}{iii} 574 line increases slightly (compared to the rate in \ciii) and 2) the rate in \ion{C}{iii} 386 shows a slightly more negative excursion at the end of the \ciii\ line formation region. The net effect is to re-populate \lsl\ and to depopulate \lsu, leading to a reduced emission. The higher rates in \ion{C}{iii} 386 and \ion{C}{iii} 574 can be understood as a desaturation effect. Due to the higher mass loss rate, the formation region is shifted towards larger velocities. The associated Doppler shift allows \ion{C}{iii} 386 and \ion{C}{iii} 574 to absorb more photons from the adjacent continuum, and consequently the radiative rates in those lines are increased. As we have seen in section \ref{form_4650}, \ion{C}{iii} 884 does not play an important role in the formation of \ion{C}{iii} 4650 in the high temperature model. 

In the 30500 K model, the increase of the emission when \mdot\ strengthens from $3 \times\ 10^{-8}$ to $2 \times\ 10^{-6}$ \myr\ is mainly due to a growing contribution of the wind to the line profile, as at higher temperature. The fraction of the line formed above 10 \kms\ grows from 0.6\%\ for \mdot\ = $3 \times\ 10^{-8}$ \myr\ to 23.5\%\ for \mdot\ = $2 \times\ 10^{-6}$ \myr. Consequently, the contribution of the wind (emission\footnote{The wind contribution is the radiation coming from material above the photosphere. Radiation is emitted in all directions in spectral lines and a fraction of the photons are parallel to the sight line, thus adding to the light directly coming from the photosphere.}) is larger for higher mass loss rates.
In addition to this dominant effect, there is also a higher radiative rate in the \ion{C}{iii} 884 line in the photosphere (below 10 \kms). In the same time, the rates in the \ion{C}{iii} 386 and \ion{C}{iii} 574 lines are not significantly modified compared to the other rates. Desaturation is responsible for the enhanced rate in \ion{C}{iii} 884. A lower \lsl\ population results from this, with the consequence of stronger \ciii\ emission. All in all, a higher mass loss rate leads to a stronger \ciii\ emission at low \teff.

Wind effects can also explain the behavior of the \ion{C}{iii} 4650 triplet in Fig.\ \ref{spec_wind}. The higher the mass loss rate, the larger the fraction of the line formed at velocities higher than 10 \kms. The wind contribution to the line profile increases, implying more emission at high effective temperature and less absorption at low temperature. In addition, at high \teff, the UV lines populating the upper and lower level of \ion{C}{iii} 4650 are formed at velocities larger in the high \mdot\ model than in the low \mdot\ model. They all have positive radiative rates over the line formation region. As a result \ion{C}{iii} 538 depopulates \ltl\ as soon as Doppler shifts allow the line to absorb more continuum photons. In the same time, \ion{C}{iii} 1256, \ion{C}{iii} 1428, \ion{C}{iii} 1578, \ion{C}{iii} 2009 populate \ltu. The effect is an increase of the \ion{C}{iii} 4650 emission, as observed in Fig.\ \ref{spec_wind}.

\end{appendix}

%%%%%%%%%%%%%%%%%%%%%%%%%%%%%%%%%%%%%%%%%%%%%%%%%%%%%%%%%%%%%%%%%%%%%%%%%%%%%%%%%%%%%%%%%%%%%%%%%%%%%%%%%%%%%%%%%%%%%%%%%%%%
%%%%%%%%%%%%%%%%%%%%%%%%%%%%%%%%%%%%%%%%%%%%%%%%%%%%%%%%%%%%%%%%%%%%%%%%%%%%%%%%%%%%%%%%%%%%%%%%%%%%%%%%%%%%%%%%%%%%%%%%%%%%
\end{document}